# Martian meteorites reflectance and implications for rover missions

L. Mandon[1*], P. Beck[2,3], C. Quantin-Nataf[1], E. Dehouck[1], A. Pommerol[4], Z. Yoldi[4], R. Cerubini[4], L. Pan[1], M. Martinot[1], V. Sautter[5]

[1] Univ Lyon, Univ Lyon 1, ENSL, CNRS, LGL-TPE, F-69622, Villeurbanne

[2] Université Grenoble-Alpes, CNRS, IPAG, UMR 5274, Grenoble, France

[3] Institut Universitaire de France

[4] Space Research & Planetary Sciences Division Physikalisches Institut, Universität Bern, Bern, Switzerland

[5] Institut de Minéralogie, de Physique des Matériaux et de Cosmochimie, Muséum National d'Histoire Naturelle, 75005, Paris, France

*Corresponding author. Now at LESIA, Observatoire de Paris, Université PSL, CNRS, Sorbonne Université, Université de Paris, 5 place Janssen, 92195 Meudon, France. E-mail address: lucia.mandon@obspm.fr

## Highlights

- We provide the first VNIR (0.4–3 µm) spectral database representative of the current Martian meteorites' diversity
- Point measurements and hyperspectral cubes are acquired on unprocessed rock samples, similarly to present and future measurements of the SuperCam and MicrOmega instruments onboard Martian rovers
- Bidirectional spectral measurements are performed on a shergottite rock and powder, in order to characterize the influence of varying illumination and observation geometries on its spectral features

## Abstract

During this decade, two rovers will characterize in situ the mineralogy of rocks on Mars, using for the first time near-infrared reflectance spectrometers: SuperCam onboard the Mars 2020 rover and MicrOmega onboard the ExoMars rover, although this technique is predominantly used in orbit for mineralogical investigations. Until successful completion of sample-return missions from Mars, Martian meteorites are currently the only samples of the red planet available for study in terrestrial laboratories and comparison with in situ data. However, the current spectral database available for these samples does not represent their diversity and consists primarily of spectra acquired on finely crushed samples, albeit grain size is known to greatly affect spectral features. Here, we measured the reflected light of a broad Martian meteorite suite as a means to catalogue and characterize their spectra between 0.4 and 3 µm. These measurements are achieved using a point spectrometer acquiring data comparable to SuperCam, and an imaging spectrometer producing hyperspectral cubes similarly to MicrOmega. Our results indicate that point spectrometry is sufficient to discriminate the different Martian meteorites families, to identify their primary petrology based on band parameters, and to detect their low content in alteration minerals. However, significant spectral mixing occurs in the point measurements, even at spot sizes down to a few millimeters, and imaging spectroscopy is needed to correctly identify the various mineral phases in the meteorites. Additional bidirectional spectral measurements on a consolidated and powdered shergottite confirm their non-Lambertian behavior, with backward and suspected forward scattering peaks. With changing observation geometry, the main absorption strengths show variations up to ~10–15%. The variation of reflectance levels is reduced for the rock surface compared to the powder. All the spectra presented are provided in the supplementary data for further comparison with in situ and orbital measurements.

## 1 Introduction

Reflectance spectroscopy is a non-destructive technique allowing the detection of various mineral phases without the need for sample preparation. Therefore, it is well suited for the characterization of precious samples such as Martian meteorites. It is also the prime technique used to remotely investigate the mineralogical composition of planetary surfaces and has shown great successes in the case of Mars (e.g., Mustard et al., 1993; Erard and Calvin, 1997; Bibring et al., 2005; Ehlmann et al., 2009; Carter et al., 2013). The Infrared Spectrometer of the SuperCam instrument and MicrOmega (respectively a point spectrometer and an imaging spectrometer) are VNIR (Visible and Near-Infrared) reflectance spectrometers used to characterize the mineralogy of the Martian surface. They are operating or will operate onboard the Mars 2020 (NASA; Wiens et al., 2017) and ExoMars rovers (ESA/Roscosmos; Bibring et al., 2017) respectively. These analyses will provide the opportunity to compare and link the spectra measured at the surface to the global spectral measurements from the orbiters, but also to the well-characterized Martian meteorites.

Until the Mars Sample Return (MSR) and the Martian Moon eXploration (MMX) campaigns are completed (i.e., 2031 at the earliest for MSR and 2029 for MMX; Usui et al., 2018; Muirhead et al., 2020), Martian meteorites are currently the only material from the Martian system readily available for laboratory studies and direct comparison with in situ and orbital measurements (even though most of them originate from the subsurface and might not be representative of the Martian regolith). However, their documented reflectance spectra were mainly acquired on powders, even though the physical state of the sample, and especially its grain size, have been shown to significantly influence both the absolute reflectance and the shape of the absorption bands (Crown and Pieters, 1987; Salisbury and Wald, 1992; Sunshine et al., 1993; Mustard and Hays, 1997). In addition, in situ measurements by the SuperCam instrument will be achieved remotely and without any sample grinding (Wiens et al., 2017). This demonstrates the need to acquire spectra spanning the diversity of Martian meteorites, especially on non-pulverized rock samples.

The documentation of multiple spectra of Martian meteorites would also provide additional constraints on the location of their ejection sites. Assessing these key locations would bring a unique opportunity to link the Martian meteorites extensive characterization in laboratories to the orbital geological study of the surface of Mars. In addition, the comparison between their absolute age and the crater densities of their ejection site would allow the calibration of the Martian chronology, in a joined effort with the sample return missions (e.g., Muirhead et al., 2020). Given their diversity of crystallization and cosmic-ray exposure ages (especially those of shergottites), they probably originate from different sites on Mars (the readers are referred to Nyquist et al., 2001 and Udry and Day, 2018 for a review on their age). As of today, some precise ejection sites have been proposed (e.g., Werner et al., 2014) but there is no real consensus on the source craters of these meteorites. Ody et al. (2015) conducted a global survey of the Martian surface using the Observatoire pour la Minéralogie, l'Eau, les Glaces et l'Activité (OMEGA) dataset and compared the reflectance of the terrains to the spectra of seven Martian meteorites, as a means to locate their ejection sites. This study revealed potential source regions with close spectral matches, but was limited by the number of available meteorite spectra (e.g., only four shergottite spectra, only one nakhlite spectrum, only one chassignite spectrum, and no spectrum of augite-rich basalt or polymict breccia specimen). A new analysis of the orbital global spectral datasets with a more extensive meteorite spectral database might narrow down these potential ejection regions.

With the exception of a few studies (e.g., Fadden and Cline, 2005; Hiroi et al., 2011; Manzarini et al., 2019), most of the reflectance spectra previously obtained in the VNIR range on Martian meteorites were acquired on powders (e.g., Sunshine et al., 1993; Bishop et al., 1998; Dyar et al., 2011; Beck et al., 2015, Filiberto et al., 2018). Spectra of the polymict breccia were previously acquired on a chip of NWA 7034 and NWA 8171 (Cannon et al., 2015; Izawa et al., 2018) and on a residue of a dry sawing of NWA 7533 (Beck et al., 2015). A reflectance survey on rock samples of Martian meteorites in the 0.25-2.5 spectral range was conducted by Hiroi et al. (2011) on eight shergottites and four nakhlites, two of which are in common with our sample suite. Similar measurements were performed by Fadden and Cline (2005) as a means to produce a framework to locate ejection regions on Mars. These studies show that VNIR reflectance can successfully be used to characterize the main mineralogy of Martian rocks and classify them relative to the families previously presented. A few additional spectra of Martian meteorites are available in the RELAB spectral library, mostly acquired on powders.

In this study, we aim at contributing to a broad Martian meteorites spectral database, by providing their spectra in the VNIR range between 0.4 and 3 µm, using instruments similar to the SuperCam and MicrOmega spectrometers. We use these analyses to characterize their spectral features in the VNIR range. Our spectral survey is conducted on a suite of samples representative of the current Martian meteorites' diversity, the majority of them being rock chips or cut sections. We present two types of VNIR spectral analyses: point spectrometry and imaging spectrometry. These techniques are closely analogous to what is and will be achieved by the reflectance spectrometers in SuperCam (acquiring point spectra) and MicrOmega (acquiring spectral cubes). In addition to the spectra of 11

Martian meteorites which were already measured in the literature mostly on particulate samples (ALH[1] 77005, ALH 84001, EETA[2] 79001, Lafayette, Los Angeles, MIL[3] 03346, Nakhla, NWA[4] 2737, NWA 7034, Tissint and Zagami), we document the reflectance spectra of 16 meteorites whose VNIR spectra have not been reported to date: NWA 480, NWA 4766, NWA 12633, NWA 12960, NWA12965 (basaltic shergottites), DaG[5] 476, DaG 489, NWA 1195 (olivine-orthopyroxene-phyric shergottites), NWA 1068, SaU[6] 008 (olivine-phyric shergottites), NWA 1950, NWA 4468, NWA 7397 (poikilitic shergottites), CeC[7] 022, NWA 817 (nakhlites) and NWA 8159 (augite-rich basalt)[8]. Eleven meteorites from our suite, including the polymict breccia NWA 7034, are analyzed with imaging spectroscopy, providing refined details on the spectral features of the different Martian meteorite families and further interpretation of the point spectra obtained.

Planetary surfaces as well as meteorites do not behave as Lambertian surfaces, meaning that reflectance is strongly influenced by the illumination and observational geometries (Hapke, 1993; Beck et al., 2012). This is expressed through anisotropy in the scattered light, depending on the optical properties of the surface, its roughness, and the size, shape and porosity of the grains. Hence, variations of the absolute measured reflectance and band depths can be observed for the same sample at different observational geometries, while these parameters are commonly used in spectra description to characterize a sample. The SuperCam instrument will achieve VNIR spectral measurements at various observational geometries due to variations in solar incidence, observational geometry and local topography, hence greatly restricting sample-to-sample comparison. To characterize the influence of the observational geometry on the light scattered by one typical Martian meteorite, we additionally measured the bidirectional reflectance spectra of the NWA 4766 sample at different observation angles, both on a particulate and on a non-crushed rock sample.

## 2 The Martian meteorites

### 2.1 Nomenclature

Except for the breccia "Black Beauty" (NWA 7034 and pairings), Martian meteorites are usually mafic to ultramafic magmatic rocks frequently showing cumulate textures. They were initially classified into three main families based on the first observed falls Shergotty, Nakhla and Chassigny: the shergottites, the nakhlites and the chassignites (SNCs), and later regrouped under the Martian denomination (Bogard and Johnson, 1983). The shergottites, the largest family in number of specimens, mainly contain pyroxene (pigeonite and augite), plagioclase and some of them, olivine. Abundant textural, mineralogical and compositional variability is observed within this family. In this study, the following classification is adopted: "basaltic", "olivine-phyric", "olivine/orthopyroxene-phyric" and "poikilitic" shergottites. Their petrology can be summarized as follows:

- the basaltic shergottites are fine-grained with doleritic texture, mainly composed of augite and pigeonite clinopyroxenes and maskelynite (shocked plagioclase), and are devoid of olivine (McSween, 1994);
- the olivine-phyric shergottites (hereafter abbreviated to ol-phyric shergottites) exhibit olivine megacrysts, orthopyroxene and chromite, enclosed in a fine-grained groundmass of pigeonite and maskelynite evidencing effusive texture (Goodrich, 2002);
- the olivine/orthopyroxene-phyric shergottites (hereafter abbreviated to ol/opx-phyric shergottites) are similar to olivine-phyric shergottites, with additional orthopyroxene megacrysts (Irving et al., 2004);
- the poikilitic shergottites were initially referred to as "lherzolitic shergottites" given their high olivine content (McSween, 1994). They exhibit coarse-grained olivine enclosed in orthopyroxene phenocrysts (i.e., a poikilitic texture), with interstitial groundmass filled by pigeonite and augite clinopyroxenes, and to a lesser extent than in basaltic shergottites, maskelynite. However, as exposed by Walton et al. (2012), they differ from the strict definition of terrestrial lherzolites in that they are not plutonic rocks, and some of them have too little olivine and too much plagioclase. Consistently, the terminology poikilitic shergottite (Walton et al., 2012) is preferred here.

---

[1] Allan Hills
[2] Elephant Moraine
[3] Miller Range
[4] Northwest Africa
[5] Dar al Gani
[6] Sayh al Uhaymir
[7] Caleta el Cobre
[8] The sample selection reflects the current collection of Martian meteorites at the IPAG (Grenoble, France) and the ENS (Lyon, France) laboratories, with additional samples lent by the University of Brest and the IPGP (Paris, France).

NWA 7635 is officially classified as a shergottite by the Meteoritical Society; it however differs from the four shergottite groups described above. It exhibits plagioclase phenocrysts shocked to maskelynite, olivine and augite, but no pigeonite or phosphate phase was observed, contrary to the other shergottites (Lapen et al., 2017). Another classification of shergottites is based on their content in light Rare Earth Elements (REE): "enriched", "intermediate" and "depleted", relatively to chondrites (e.g., Barrat et al., 2002; Borg et al., 2003; Basu Sarbadhikari et al., 2009). The second most abundant SNC family, the nakhlites, consist of clinopyroxene cumulates. They are augite- and to a lesser extent olivine-rich (for a review, see Treiman, 2005). Only three occurrences of chassignites, the dunite type of SNC, were reported to date. In addition to the SNCs, three other meteorites were classified as Martian: a polymict breccia (the biggest specimen being NWA 7034, Agee et al., 2013; Humayun et al., 2013; Wittmann et al., 2015), the orthopyroxenite ALH 84001 (Mittlefehldt, 1994) and the augite-rich basalt NWA 8159 (where pigeonite is absent unlike in shergottites, Ruzicka et al., 2015). The polymict breccia NWA 7034 and pairings is a unique type of meteorite consisting of various lithic (norite, monzonite, gabbro), mineral (feldspar, pyroxene, iron-oxides, apatite) and melt (rock and spherule) clasts, enclosed in a fine-grained matrix enriched in magnetite and maghemite oxides (Agee et al., 2013; Humayun et al., 2013).

Over the last decades, controversy has arisen on whether the shergottites are young (<1 Ga) or old (>4 Ga) Martian rocks. On the one hand, analyses of individual minerals with the $^{87}$Rb–$^{87}$Sr, $^{147}$Sm–$^{143}$Nd, $^{176}$Lu–$^{176}$Hf and U–Pb chronology systems yield fairly young radiometric ages for the shergottites (175–475 Ma; for a review, see Nyquist et al., 2001), interpreted as a crystallization during the Amazonian era on Mars. On the other hand, bulk-rock analyses on the $^{207}$Pb–$^{206}$Pb and $^{87}$Rb–$^{87}$Sr systems return contrasting radiometric ages at ~4.1 Ga (Bouvier et al., 2008). The shergottites' cosmic ray exposure age points to ejections from Mars about ~0.5 Ma to ~20 Ma ago (Bogard et al., 1984; Shukolyukov et al., 2002). Unlike the shergottites, there is little debate on the age of nakhlites and chassignites, which share similar ejection and crystallization ages (~11 Ma and ~1.3 Ga respectively), pointing to a single ejection event and a common origin on Mars (Bogard et al., 1984; for a review on formation ages, see Borg and Drake, 2005). The oldest specimen is the orthopyroxenite ALH 84001, whose ~4.5 Ga age yields a formation during the very early period of Mars (Nyquist et al., 1995). Additionally, some zircons in the polymict breccia NWA 7533 have ages of ~4.4 Ga (Humayun et al., 2013).

**2.2 Alteration minerals in Martian meteorites**

On Mars, orbital data as well as in situ measurements have shown the presence of various alteration minerals: Fe/Mg and Al-smectites, kaolins, micas, serpentines, chlorites, prehnites, sulfates, carbonates, zeolites, oxides and hydrated silica (Christensen et al., 2000; Bibring et al., 2005; Gendrin et al., 2005; Ehlmann et al., 2008; Ehlmann et al., 2009; Carter et al., 2013; Vaniman et al., 2014). Martian meteorites are predominantly composed of pyroxene, olivine and plagioclase (mainly in the form of maskelynite), with minor accessory phases such as phosphates (apatite, merrillite). These primary minerals are sometimes associated with alteration phases, of preterrestrial origin (formed on Mars or during transport to the Earth) or produced during terrestrial residence. The latter occurs especially in hot desert find meteorites, which were largely the focus of this study (e.g., Jull et al., 1990; Scherer et al., 1992). Hence, it is likely that terrestrial weathering might play a significant role in forming any secondary phase detected in the VNIR spectra presented in this study. The meteorites found in hot deserts are usually more fractured than cold desert finds and can exhibit partial dissolution of primary minerals and formation of secondary phases such as clay minerals, Fe-oxides/(oxy)hydroxides, carbonates, sulfates and silica (Scherer et al., 1992; Lee and Bland, 2004).

In nakhlites, secondary minerals were reported in association with olivine and in the mesostasis. These assemblages consist of Fe/Mg/Al-silicate amorphous phases, poorly-crystalline phyllosilicate mixtures (likely containing smectite and serpentine), various carbonates, sulfates, salts, laihunite, ferrihydrite and iron oxides (Gooding et al., 1991; McCubbin et al. 2009; Noguchi et al., 2009; Changela and Bridges, 2010). Ca and Mg-carbonates as well as Ca-sulfates are also present in chassignites and shergottites, but phyllosilicates rarely occur (Gooding et al., 1988; Gooding et al., 1991; Wentworth and Gooding, 1994; Wentworth et al., 2005). In ALH 84001, corroded pyroxenes are associated with Ca, Fe and Mg-carbonates in concentric disks, referred to as "rosettes" (Mittlefehldt, 1994). Except from ALH 84001, most of the carbonate assemblages in Martian meteorites occur in veins and are mostly terrestrial alteration products (Lee and Bland, 2004; Gillet et al., 2005; Treiman et al., 2007; Howart et al., 2014, Shearer et al., 2015). For example, Figure 1 shows a light, likely carbonate-filled veins spreading through the slab of shergottite NWA 7397 and NWA 1068 (2). Various oxides are reported in all Martian meteorites families (such as ilmenite, ulvöspinel, chromite and magnetite; for a review, see Papike et al., 2009). In particular, the polymict breccia NWA 7034 and pairs as well as the augite-rich basalt NWA 8159 exhibit elevated content in iron oxides magnetite and maghemite (respectively ~15 wt% and ~6 wt%, Gattacceca et al., 2014; Herd et al., 2017). Similarly, the unusual dark color of the olivine grains in NWA 2737 is attributed to the presence of nanophase kamacite alloy particles (Pieters et al., 2008).

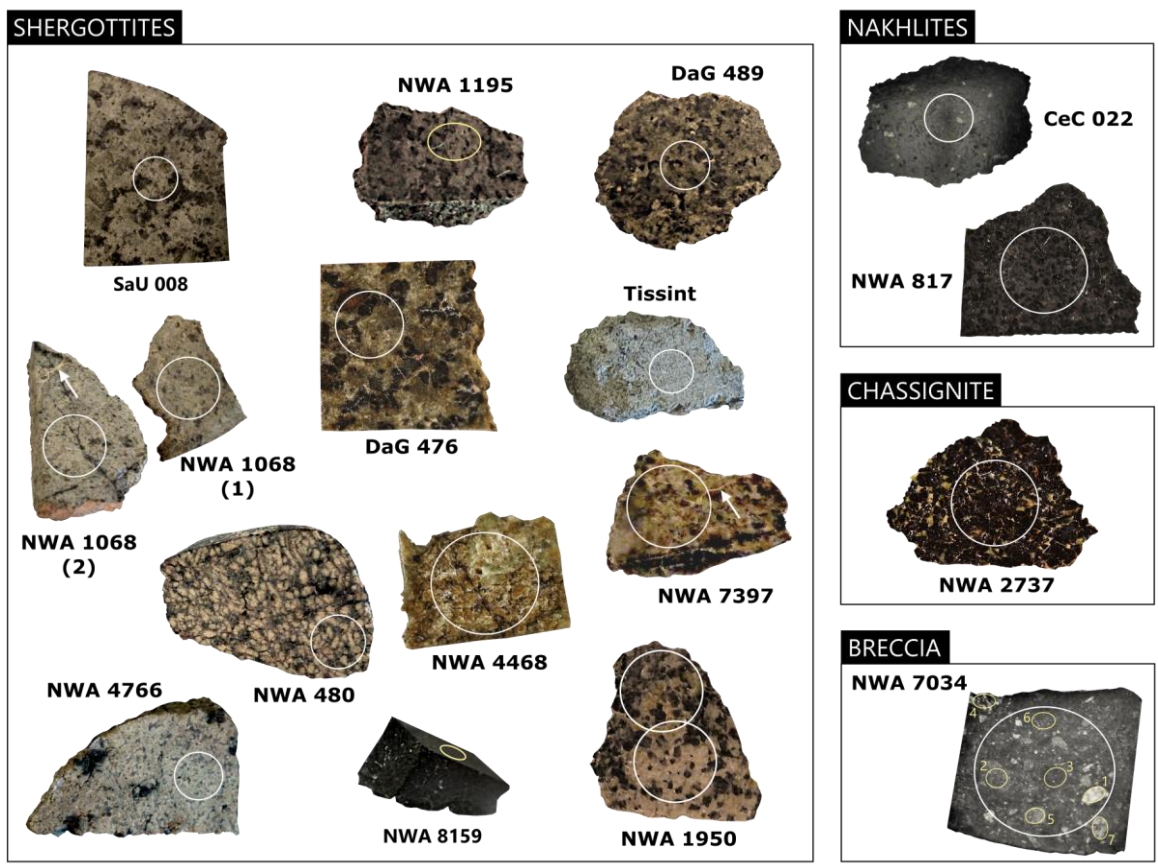

**Figure 1.** Martian meteorites non crushed rock suite and location of spectral measurements. The scale is given by the size of the measurement spot: white spots are ~5.2 mm large (regular measurement) and yellow spots are ~1.3 mm by ~1.7 mm. Two spectra were obtained for NWA 1950, which exhibit a poikilitic (light region) and a non-poikilitic area (dark region; Gillet et al., 2005). Arrows indicate the carbonate-filled veins described in the text.

## 2.3 Absorption bands of the minerals of interest (in the 0.4-3 µm region)

*2.3.1 Absorption bands of primary minerals in Martian meteorites*

Martian meteorites are mostly composed of the mafic minerals olivine and pyroxene, as well as some shocked plagioclase in maskelynite form. These minerals present electronic absorptions in the 0.4–3 µm range caused by the presence of transition metals, the most intense absorptions being related to iron, especially when in ferrous state ($Fe^{2+}$).

Pyroxenes (($Ca,Mg,Fe)_2Si_2O_6$) have reflectance spectra that exhibit two main absorption bands: around ~1 µm and ~2 µm, hereafter referred to as bands I and II (Adams, 1974; Fig. 2). The exact center positions of these absorptions are greatly modulated by the substitution of cations in the crystal field, here Fe, Mg and Ca (Adams, 1974; Cloutis and Gaffey, 1991; Klima et al., 2011). In clinopyroxenes (Cpx) spectra, which show a wide variability in Ca content, band I and band II shift to longer wavelengths with increasing Ca, with the shift of band I being subtler than that of band II (Adams et al., 1974; Fig. 2). Among the major types of Cpx present in Martian meteorites, pigeonite is Ca-poor and has band I and II usually positioned before 1 and 2 µm respectively, while augite is Ca-rich and has band I and II usually positioned after 1 µm and 2.2 µm respectively (Fig. 2). Orthopyroxenes (Opx) are depleted in Ca: their band I and II are usually positioned before 1 and 2 µm. Within a pyroxene family with fixed Ca content such as the orthopyroxenes, the band center shift is mainly modulated by the Fe content, with absorptions occurring at longer wavelength with increasing Fe (Cloutis and Gaffey, 1991).

The main absorption of olivine (($Mg,Fe)_2SiO_4$) is located in the ~0.7-1.8 µm range and centered around ~1 µm (Fig. 2). The shape of the absorption is modulated by the Fe over Mg content: fayalite, the Fe-endmember of olivine exhibits a wider and more flattened band than forsterite, the Mg-endmember (King and Ridley, 1987; Fig. 2).

Visible and near-infrared spectra of plagioclase feldspar (($Na,Ca)(Si,Al)_4O_8$), when Fe-bearing, are characterized by a weak absorption in the 0.8-1.7 µm range and centered around ~1.1-1.3 µm, deepening with increasing Fe (Adams and Goullaud, 1978; Fig. 2). Of relevance for maskelynite identification in reflectance data, shocked plagioclase experiments show a drop of the absolute reflectance and a decrease in the absorption band strength with increasing shock pressures (Johnson and Hörz, 2003). Mixed with pyroxene or olivine, plagioclase is usually detected only if present in high abundance and rich in iron (the limit of detection depends on the grain size; Crown and Pieters, 1987).

The iron and nickel alloys detected in Martian meteorites absorb efficiently at VNIR wavelengths and usually show a red slope in this range (Fig. 2). Finally, phosphates can exhibit various absorption bands related to Fe electronic processes if iron-bearing, and vibrations of the $PO_4^{3-}$, $H_2O$ and OH groups if hydrated or hydroxylated. Merrillite and apatite are depleted in iron; hence, if neither hydrated nor hydroxylated, their spectra are relatively featureless, the $PO_4^{3-}$ bands being usually very subtle (for a review, see Bishop et al., 2019; Fig. 2).

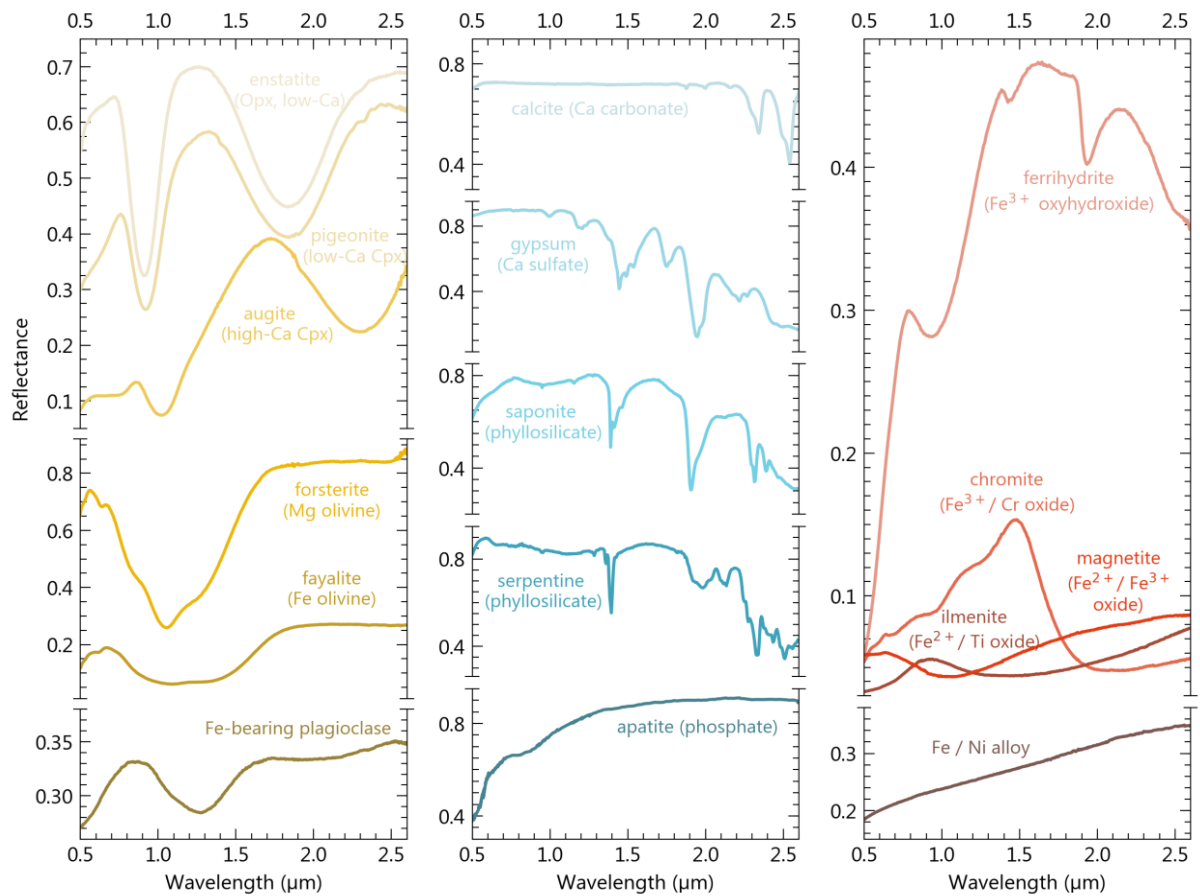

**Figure 2.** Laboratory reflectance spectra of various minerals reported in Martian meteorites and absorbing in the VNIR range, mainly taken from the PDS Geoscience Spectral Library. Spectra IDs are as follows: enstatite C2PE30, augite C1SC35, pigeonite C1PP42, forsterite C3PO51, fayalite C3PO59, plagioclase NCLS04, calcite CBCB12, gypsum F1CC16B, saponite LASA51, serpentine LASR10, ferrihydrite C1JB45, chromite LACR12, magnetite LAMG07, ilmenite LAPI06A and Fe/Ni alloy C1SC12. Apatite spectrum is taken from Lane et al. (2007). The absolute reflectance values are conserved in the plots; however, we encourage the reader to keep in mind that these values are dependent from the sample grain size adopted in the spectral measurement (e.g., Salisbury and Wald, 1992; Mustard and Hays, 1997).

*2.3.2 Absorption bands of alteration minerals in Martian meteorites*

Infrared spectrometry is a powerful tool to assess the presence of alteration minerals. In particular, combination of bending and stretching of groups such as $H_2O$, $OH^-$, metal–OH, $CO_3^{2-}$ and $SO_4^{2-}$ in minerals are responsible for the majority of the vibrational absorptions observed in the near-infrared for the alteration minerals reported in Martian meteorites. The water molecule and the hydroxyl ion produce an intense absorption band centered at 3 µm as well as two narrow bands at 1.4 µm and 1.9 µm (Hunt, 1977; Fig. 2). These two absorptions usually require the presence of water at a structural level in minerals to appear, while the 3 µm band is commonly observed in laboratory spectral measurements and does not require large amounts of water to appear (e.g., if adsorbed water is present on the sample). Also of relevance for the hydrous minerals detection, hydroxyl bounded to metal cations (such as $Fe^{2+/3+}$, $Mg^{2+}$, $Al^{3+}$, $Si^{4+}$, etc.), creates additional narrow absorptions in the near-infrared (Hunt, 1977; Fig. 2). The phyllosilicate assemblages reported in nakhlites (i.e., smectite and serpentine) present these kinds of absorptions (Fig. 2). Carbonates have specific absorptions related to the $CO_3^{2-}$ ion, which appear in the 0.4-3 µm region at ~2.3 µm and ~2.5 µm (Hunt and Salisbury, 1971; Fig. 2). Sulfate minerals have additional specific absorptions related to the $SO_4^{2-}$ ion, mostly between ~2.1 µm and 2.7 µm (Cloutis et al., 2006; Fig. 2). Finally, the iron-oxides occurring in meteorites present various broad absorptions in the VNIR resulting from the electronic processes related to iron (Sherman et al., 1982; Fig. 2). These oxides are very efficient at absorbing the VNIR light, with absolute reflectance commonly below 15% (Fig. 2). As a matter of fact, the unique darkness of the polymict breccia NWA 7533 has been attributed to its elevated magnetite and maghemite contents (Beck et al., 2015).

## 3 Samples and methods

### 3.1 Meteorite suite

We gathered a sample suite representative of the Martian meteorite's diversity described in section 2.1 (Fig. 1 and 3). This suite comprises:
- The orthopyroxenite ALH 84001;
- The polymict breccia NWA 7034;
- The augite-rich basalt NWA 8159;
- One out of the 3 chassignites NWA 2737;
- Five out of the 13 nakhlite pairings[9] (see Fig. 3 for sample names);
- 18 shergottites out of the 204 individual specimens (with likely pairings) approved by the Meteoritical Society, including all the main classes (see Fig. 3 for sample names):
    o Seven basaltic shergottites;
    o Three olivine/orthopyroxene-phyric shergottites (including at least a pairing);
    o Four olivine-phyric shergottites;
    o Four poikilitic shergottites.

The locations of the spectral measurements on the rock samples are shown in Figure 1.

Figure 3 shows the modal mineralogy of these meteorites (values retrieved from the literature). Consistently with the Martian meteorite global diversity and except some effusive ol/opx- and ol-phyric shergottites, most of our samples are consistent with cumulates hypovolcanic rocks (i.e., excavated from depth), as indicated by the predominance of pyroxene and olivine. Here, the only meteorite that exhibits a modal mineralogy consistent with a surface basalt resulting from effusive processes from a non-fractioned primary melt is Los Angeles.

---

[9] As of 2020

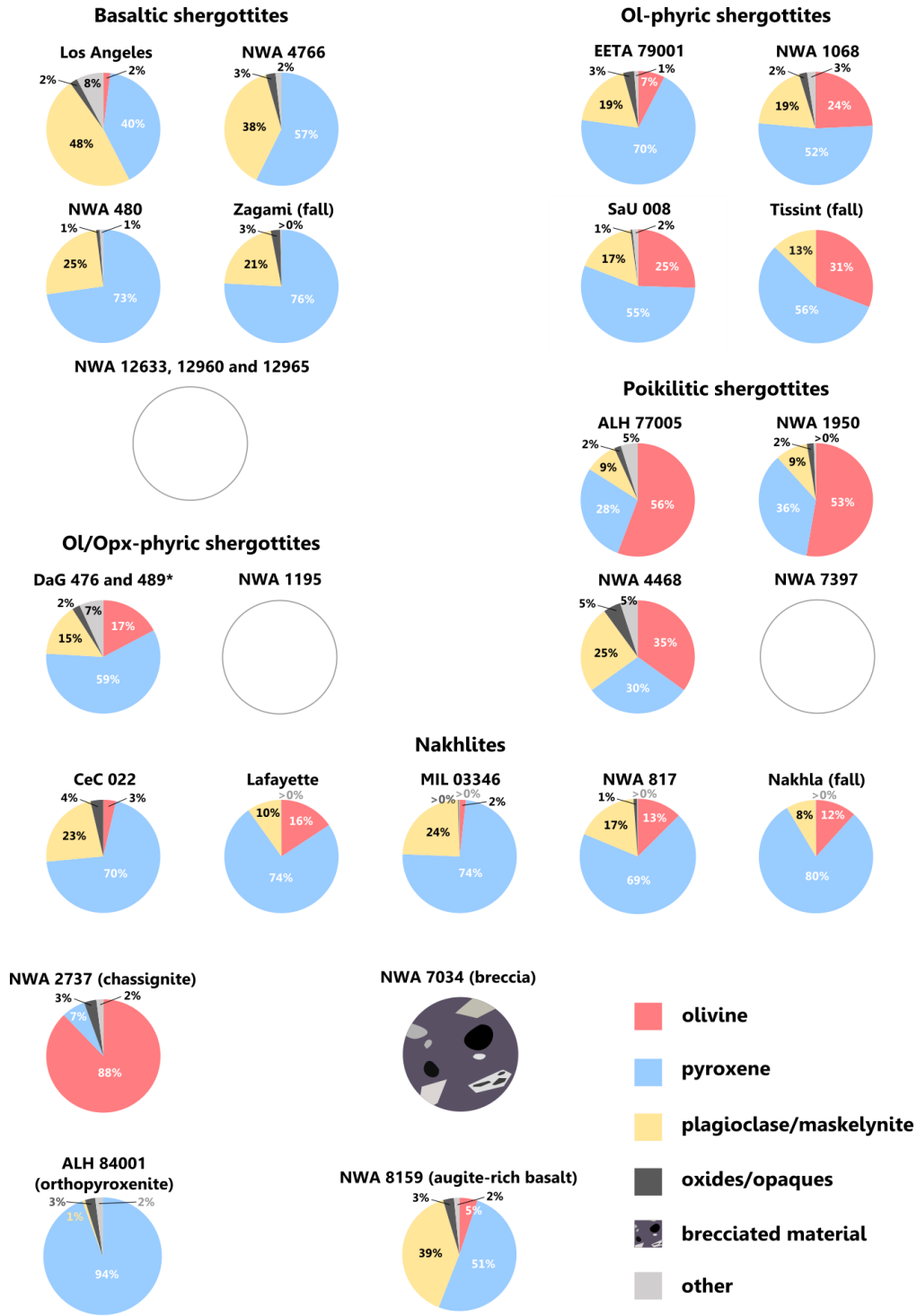

Fig 3. Classification of the Martian meteorites analyzed and modal mineralogy retrieved from the literature (averaged from Stolper & McSween 1979; Ma et al., 1981; Mason 1981; Treiman et al., 1994; Gleason et al., 1997; Lentz et al., 1999; Folco et al., 2000; Rubin et al., 2000; Zipfel et al., 2000; Mikouchi 2001; Wadhwa et al., 2001; Barrat et al., 2002; Gillet et al., 2002; Mikouchi & Miyamoto 2002; Sautter et al., 2002; Xirouchakis et al., 2002; Bartoschewitz et al., 2003; Russell et al., 2004; Anand et al., 2005; Gillet et al., 2005; McKay & Schwandt 2005; Mikouchi et al., 2005; Rutherford et al., 2005; Stopar et al., 2005; Beck et al., 2006; Day et al., 2006; Wittke et al., 2006; Imae & Ikeda 2007; Irving et al., 2007; Treiman et al., 2007; Walton and herd 2007; Lapen et al., 2010; Hsu et al., 2012; Ruzicka et al., 2015; Herd et al., 2017; Krämer Ruggiu et al., 2019). Empty circles correspond to lack of values.

* DaG 476 and DaG 489 are likely paired together (Grossman, 1999).

Some of the cut section samples were covered by resin in the past, which was removed by thin polishing of the sections by previous owners. As the resin percolated through the grains of the samples, a removal of this leftover resin without a destructive method could not be achieved. These residues can appear in our spectra as contaminants, but their spectral signatures do not affect the major mineral bands and are not an issue for identifying the main mineralogy in the point spectra and hyperspectral cubes. However, some types of resin can exhibit absorption bands similar to those of secondary minerals, at 1.9 µm or 2.3 µm and 2.5 µm (e.g., Li et al., 2017). This issue was considered when interpreting the various point spectra and hyperspectral cubes of cut sections mounted in resin.

### 3.2 Point spectra acquisition

The point reflectance spectra of the meteorites were acquired using the Spectrophotometer with cHanging Angles for the Detection Of Weak Signals (SHADOWS, Potin et al., 2018) designed at the Institut de Planétologie et d'Astrophysique de Grenoble (IPAG). SHADOWS is a spectro-goniometer capable of acquiring individual spectra of dark samples (>0.005% in reflectance) at various geometries of illumination and observation angles (Potin et al., 2018). The light source is located on the first arm of the goniometer and consists of a monochromatic beam transported through eight optical fibers and focused on the sample. The reflected light is collected by two detectors, one covering the VNIR range from 0.185 µm to 1.2 µm, the other covering the infrared range from 0.8 µm to 5.2 µm, and located on a second arm (Potin et al., 2018). The Martian meteorites spectra were measured at a spectral sampling of 20 nm and a variable spectral resolution (on the order of a few tens of nanometers, Potin et al., 2018) mainly between 0.36 µm and 3 µm, as this range contains the majority of absorption bands of the minerals of interest (both primary and secondary phases; e.g., Singer et al., 1981; Ody et al., 2012; Carter et al., 2013; Fig. 2).

*3.2.1 Primary spectral measurements*

We performed measurements on all the samples at a phase angle $\phi$ = 30°, corresponding to an incidence angle $\theta_i$ of 0° (illumination at nadir) and an emergence (observation) angle $\theta_e$ of 30°. The incidence and emergence angles $\theta_i$ and $\theta_e$ are here defined as the angle between the normal to the surface and the incident beam, and between the normal to the surface and the observed direction, respectively. The phase angle $\phi$ corresponds to the angle between these two directions. At the geometry of measurement, the resulting illumination spot size is ~5.2 mm wide (Potin et al., 2018; Fig. 1). For smaller samples, a pinhole of 500 µm is placed to select the incident beam of only one optical fiber while masking the seven others, resulting in a ~1.3 mm by ~1.7 mm illumination spot and lower signal-to-noise (SNR) ratio (Potin et al., 2018; Fig. 1). For each day of measurement, spectra of reference targets (Spectralon™ and Infragold™) are acquired and used to calibrate the data. Additional software correction of the data also includes adjustments related to the photometric angle used in the measurements (e.g., modification of the illumination spot shape, non-Lambertian behavior of the spectralon, see Potin et al., 2018).

*3.2.2 Spectral measurements with varying observation geometry*

We characterized the anisotropy of the light scattered by NWA 4766, both on powder and cut section, by acquiring bidirectional spectral measurements between 0.36 µm and 3 µm with the point spectrometer SHADOWS, at various incidence, emergence and azimuth angles. This sample was selected because it is part of an abundant type of Martian meteorites, the basaltic shergottites (see section 2.1), and because it shows no obvious indication of terrestrial alteration (e.g., no vein crosscutting the melt veins and pockets, no absorption band related to the presence of secondary minerals in the sample). The powder was prepared by crushing a few milligrams of NWA 4766 in an agate mortar. Similar to the sample preparation detailed in Beck et al. (2012), the powder was not sieved, as a means to produce a particulate sample similar to a natural particulate medium, with various grain sizes (on the order of tens of microns maximum). To ensure accurate similitude of the measurements on the rock and the powder, homogeneous areas of the meteorite were considered for the spectra acquisition on the chip, and for sampling of the area to be crushed for powder preparation.

Measurements at positive azimuth angles (outside of the principal plane) were performed by rotating the observation arm in the horizontal plane. High phase angle measurements are limited by the extent of the measured surface and consequently the quantity of available meteorite material (i.e., high phase angle measurements require a wide surface as a result of the illumination spot elongation). No measurement was made at phase angles lower than 10°, as in this configuration, parts of the goniometer illumination arm are masking the illumination spot from the detector. The geometries of measurements explored are given in Table 1.

**Table 1**. List of observational geometries used for bidirectional spectral measurements on Martian meteorite NWA 4766 (basaltic shergottite).

| Sample | Incidence $\theta_i$ | Emergence $\theta_e$ | Azimuth a | Min. phase angle $\phi$ | Max. phase angle $\phi$ |
| --- | --- | --- | --- | --- | --- |
| NWA 4766 (powder) | 0°, 20°, 40°, 60° | -60°, -50°, -40°, -30°, -20°, -10°, 0°, 10°, 20°, 30°, 40°, 50°, 60° | 0° | 10° | 120° |
| NWA 4766 (cut section) | 0°, 40° | -50°, -40°, -30°, -20°, -10°, 0°, 10°, 20°, 30°, 40°, 50° | 0°, 44° | 10° | 90° |

**3.3 Band parametrization**

To fully describe the absorption bands of each meteorite's point spectrum, we use the following parameters: band position center, band width, band Full Width at Half Maximum (FWHM) and band strength (or band depth, abbreviated BD). In order to retrieve these band parameters, we first perform continuum removal, using the following procedure:

- For spectra showing only the two main absorption bands of pyroxene at ~1 μm and ~2 μm, a spectral continuum is estimated for each of these two bands, similarly to the method used by Horgan et al. (2014) and Martinot et al. (2018). For each band, a set of potential linear continuums defined by a series of potential tie points are computed: the optimized continuum is the one which maximizes the area of the absorption band (Fig. 4a). For the band centered at ~1 μm, left shoulder tie points are set between 0.6 μm and 0.9 μm and right shoulder between 1 μm and 1.5 μm. For the band centered at ~2 μm, left shoulder tie points are set between 1.1 μm and 1.9 μm and right shoulder between 2.2 μm and 2.7 μm. Visual examination of the best-modelled continuum is performed, and the tie points values are adjusted if required. The continuum-removed bands are then computed by dividing bands I and II by their respective modelled linear continuum (Fig. 4c).
- For spectra with narrow absorption bands but no broad bands related to mafic minerals, an upper convex hull is computed and divided from the original spectrum. If needed, this process is repeated until all the absorption bands are correctly wrapped in the hull.
- In the case of spectra exhibiting narrow bands superimposed on the broad bands of mafic minerals, a first continuum-removed spectrum (CRS) is obtained by maximizing the pyroxene band area (first method described above). The two bands of pyroxene without the superimposed bands are modelled by removing parts of the original spectra where narrow bands are observed, and filling these parts using cubic interpolation (Fig. 4b). The resulting spectrum is used to retrieve parameters of the bands corresponding to pyroxene. In a second step, the modelled broad bands are divided from the original spectrum to retrieve the narrow superimposed absorption bands (Fig. 4d).

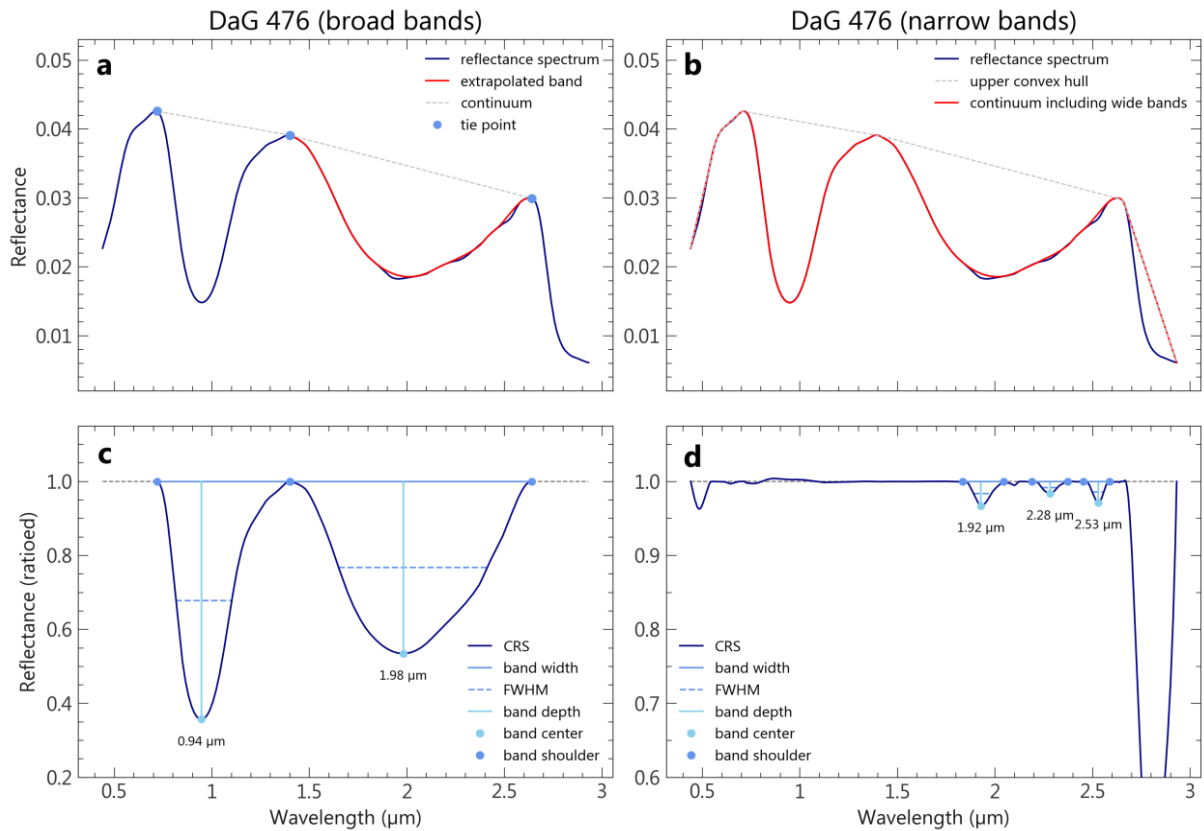

**Figure 4.** Spectral analysis performed to retrieve band parameters, illustrated for Martian meteorite DaG 476. **(a)** The continuum line is estimated by maximizing each band area between variable tie points. When narrow absorption bands are superimposed to a broad band of pyroxene, the corresponding broad band is extrapolated above the narrow bands. **(b)** When narrow absorptions are present, an additional continuum which is composed of the broad bands and an upper convex hull outside of the broad bands is computed. By doing so, both bands superimposed to the broad bands and bands located outside of the broad bands can be correctly detected. **(c)** Parametrization of the wide absorptions related to pyroxene. The CRS is obtained by dividing the original spectrum of DaG 476 corrected from the presence of the narrow absorption bands by the continuum shown in (a). **(d)** Parametrization of the narrow bands related to secondary minerals. The slight convexity in the spectrum causing the detection of a band at 0.48 µm is not considered in this study, hence the band was manually ignored. The CRS is obtained by dividing the original spectrum of DaG 476 by the continuum including the wide absorptions of pyroxene shown in (b).

Band parameters can be derived from the CRS if the left and right shoulders of the absorptions are known. For the two main bands of pyroxene, they correspond to the tie points of the optimized linear continuum (Fig. 4a and 4b). Aside from this case, a band is defined if a local minimum is found on the continuum-removed spectrum; the band shoulders are defined as the local maxima closest to this point. To exclude artificial band detection caused by noise, a threshold on the minimum absorption strength is set (usually 0.5% below the continuum baseline and adjusted if needed). Finally, the absorption bands are parameterized as follows (Fig. 4c and d):

- The band position, or band center is expressed by the wavelength at which the minimum CRS value occur, in the spectral range bounded by the band shoulders;
- The band depth is equal to 1 minus the CRS value at the band center;
- The width is defined as the right shoulder position minus the left shoulder position (in microns);
- The FWHM corresponds the width of the band at half of the band depth.

**3.4 Hyperspectral cubes acquisition**

In addition to point spectrometry, we analyzed some of the meteorite samples using imaging spectrometry, where hyperspectral images of the samples, or cubes, are produced, with the third dimension corresponding to the wavelength axis. We imaged DaG 476, DaG 489, NWA 480, NWA 1068, NWA 1195, NWA 1950, NWA 2737, NWA 4468 and NWA 7034 using a HySpex SWIR-384 imaging spectrometer. The unprocessed samples are placed on a translating table, allowing line-by-line scanning of the whole spectrum, and illuminated by a 0.4–2.5 µm broadband halogen light source at an angle of ~30° as they pass through the field of view of the detector. Hyperspectral images are acquired in 288 bands sampled every 5.45 nm and ranging from 0.93 to 2.50 µm, with a spatial resolution of ~50 µm/pixel. Additional cubes in the near-infrared were acquired on SaU 008 and Tissint at the University of Bern with the imaging system of the SCITEAS facility (Pommerol et al., 2015). Here, a narrow bandpass of light is selected by a monochromator coupled to a halogen source and illuminates the samples, while the reflected photons are collected by a camera placed at nadir and covering the near-infrared range between 0.82 and 2.5 µm. Unlike the HySpex spectrometer, a complete image of the whole sample is produced for every time step, one wavelength at a time. The light's wavelength was changed between each image using 6 nm steps, resulting in hyperspectral cubes similar to the HySpex products in terms of spectral range and sampling. The spatial resolution (~0.5 mm/px) is lower than for the HySpex measurements but remains sufficient to resolve the main phenocrysts of Tissint and SaU 008 meteorites. Internal relative calibration for both experiments is achieved using a Spectralon™ reflectance target.

To guide the spectral analysis and interpretation of the cubes, spectral index maps are produced based on the spectral features of the main absorbing minerals present in meteorites: olivine, low-calcium pyroxene (LCP, e.g., orthopyroxene and pigeonite) and high-calcium pyroxene (HCP, e.g., augite). Olivine is mapped using the olivine index described in Mandon et al. (2020): the depth of four absorption bands at 1.08 µm, 1.26 µm, 1.37 µm and 1.47 µm are measured, as well as the convexity at 1.3 µm (Mg-olivine) and 1.4 µm (Fe-olivine and coarse-grained Mg-olivine). If three out of the four absorption bands and one out of the two convexities are positive, the value of the olivine parameter corresponds to the average of the band depths. The two types of pyroxenes are mapped using an adaptation of the "*LCPINDEX2*" and "*HCPINDEX2*" parameters from Viviano-Beck et al. (2014), which are weighted averages of the depths of selected bands between 1.69 µm and 1.87 µm for the LCP index, and between 2.12 µm and 2.46 µm for the HCP index.

# 4 Spectral features of Martian meteorites

## 4.1 Results from hyperspectral cubes

False color composites in the near-infrared, parameters maps and spectra extracted from the hyperspectral cubes are shown in Figures 5 to 9. Olivine, LCP and HCP are easily identifiable both in the HySpex and SCITEAS infrared imaging products and they respectively appear blue, red/pink and green/yellow in our false color RGB composite (R = 1.22 µm, G = 1.66 µm, B = 2.15 µm). Spectral parameter maps are also successful at identifying these minerals, as confirmed by the various spectra extracted from the cubes (Fig. 5 to 8). Augite and pigeonite individual crystals or zoning are resolved for the basaltic and poikilitic shergottites, chassignite NWA 2737 and breccia NWA 7034 (Fig. 5 to 9). Even at the high spatial resolution of HySpex products, the 1.3 µm absorption feature in Fe-bearing plagioclase or maskelynite is not definitively detected, except for the basaltic shergottite NWA 480 where it might occur as bright patches in the groundmass, in association to HCP (Fig. 5b). The ol-phyric and ol/opx-phyric shergottites exhibit dark regions of recrystallized impact melt pockets, characterized by generally flat spectra and reflectance values lower than 2% in the VNIR (Fig. 6). The olivine crystals in NWA 2737 that appear extremely dark in the visible (Fig. 1) also exhibit very low reflectance in the near-infrared and a moderate red slope (Fig. 8). This is in agreement with previous studies proposing the presence of Fe/Ni alloy in olivine crystals to be responsible for their unusual dark color (e.g., Van de Moortèle et al., 2007; Pieters et al., 2008).

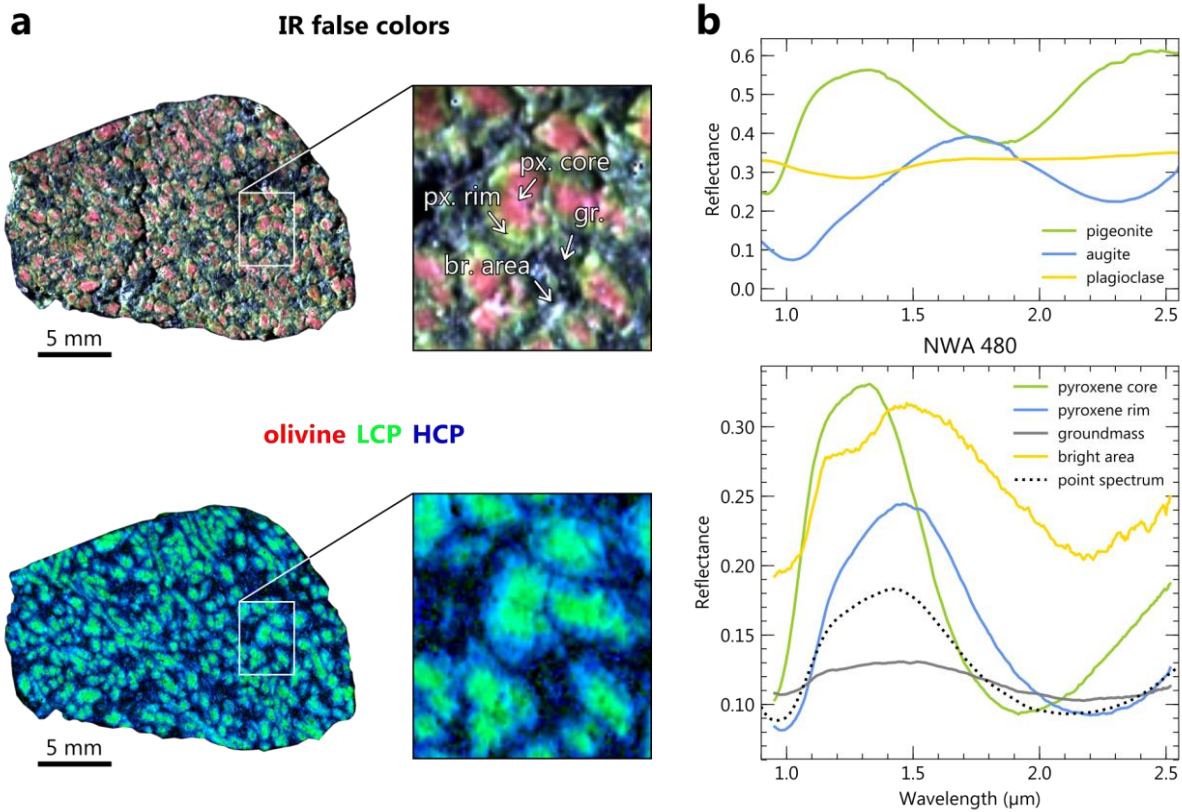

**Figure 5.** Hyperspectral cube of the basaltic shergottite NWA 480. **(a)** False RGB color composite in the near-infrared (R = 1.22 µm, G = 1.66 µm, B = 2.15 µm). In this composition, mineral zoning from LCP (pink) to HCP (green) is clearly visible. This is confirmed through examination of the RGB spectral parameters map below, where R = olivine index, G = LCP index and B = HCP index, and examination of the individual spectra. **(b)** Spectra extracted from the hyperspectral cube of NWA 480 (solid lines, location of spectral extraction shown with arrows on panel (a)) and point spectra obtained with the point spectrometer on a large part of the sample surface (dotted lines, see Fig. 1 for measurement location), compared with laboratory reflectance spectra of known minerals taken from the PDS Geoscience Spectral Library. Laboratory spectra IDs are as follows: pigeonite C1PP42, augite C1SC35 and plagioclase NCLS04. The pyroxene core and pyroxene rim spectra are consistent with LCP (represented by pigeonite) and HCP (represented by augite), respectively. The groundmass, though less reflective, has a spectrum with band centers roughly intermediate between those of low and high-calcium pyroxenes. The bright area spectrum resembles that of the groundmass, with increased reflectance level and an additional absorption at ~1.3 µm consistent with a mixture with plagioclase.

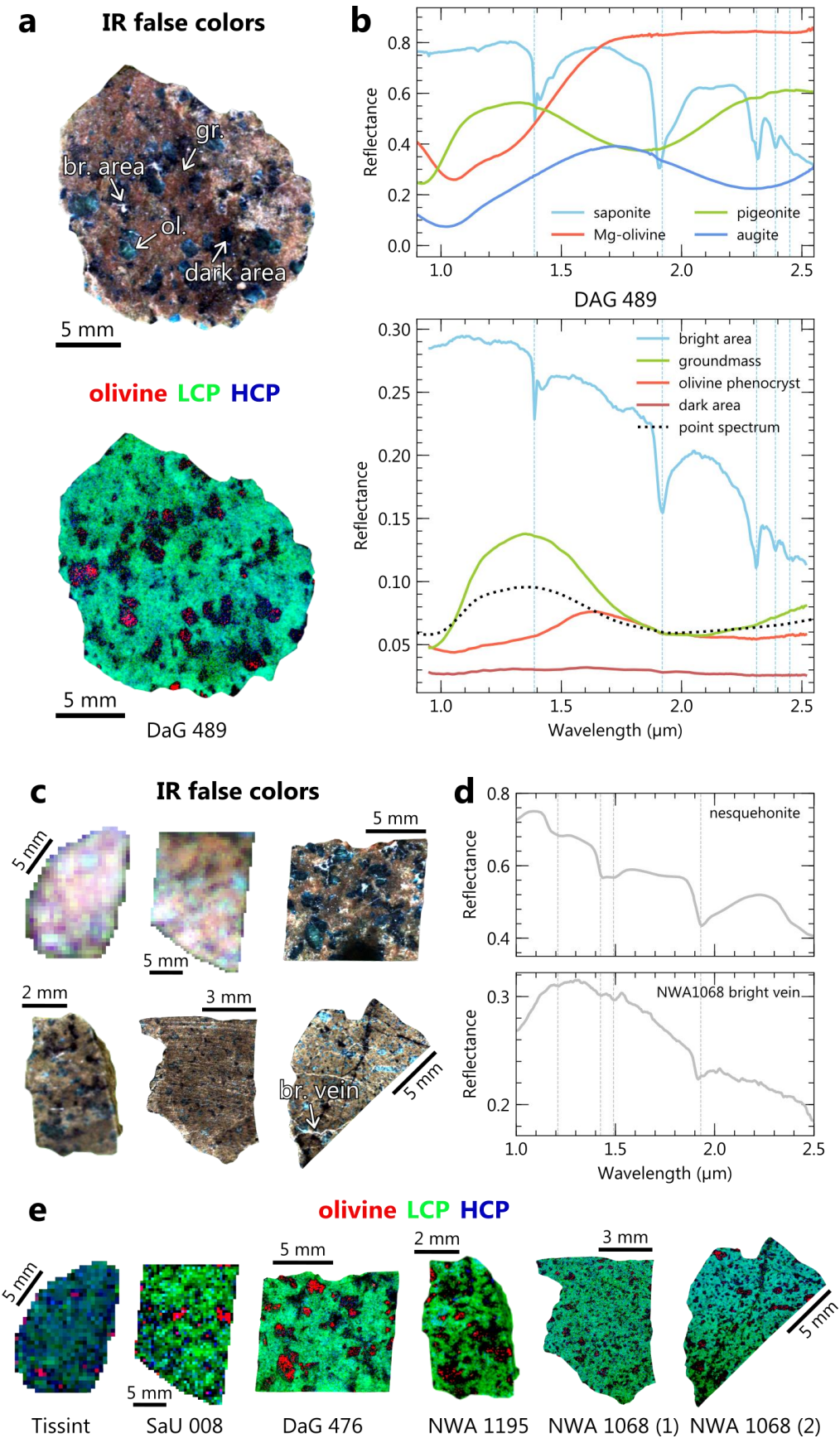

**Figure 6.** Hyperspectral cube of ol/opx-phyric and ol-phyric shergottites. **(a)** False RGB color composite of DaG 489 in the near-infrared (R = 1.22 µm, G = 1.66 µm, B = 2.15 µm) and associated RGB spectral parameters map below

(R = olivine index, G = LCP index, B = HCP index). In the IR false color composite chosen, olivine, LCP and HCP appear blue, red/pink and green/yellow respectively. **(b)** Spectra extracted from the hyperspectral cube of DaG 489 (solid lines, location of spectral extraction shown with arrows on panel (a)) and point spectra obtained with the point spectrometer on a large part of the sample surface (dotted lines, see Fig. 1 for measurement location), compared with laboratory reflectance spectra of known minerals taken from the PDS Geoscience Spectral Library. Laboratory spectra IDs are as follows: saponite LASA51, forsterite C3PO51, pigeonite C1PP42 and augite C1SC35. DaG 489 olivine phenocrysts have spectra typical of Mg-olivine. The groundmass spectrum exhibits absorption centers intermediate between low and high-calcium pyroxenes, closer to the LCP endmember. The bright patches filling the groundmass and phenocryst voids have spectra consistent with a Mg-smectite (best matched by saponite smectite). **(c)** False RGB color composite of other ol/opx-phyric and Ol-phyric shergottites in the near-infrared (R = 1.22 µm, G = 1.66 µm, B = 2.15 µm). **(d)** Spectrum extracted from a bright vein of NWA 1068 (2), compared with a spectrum of nesquehonite (hydrous Mg-carbonate) measured by Harner and Gilmore (2015). **(e)** RGB spectral parameters map (R = olivine index, G = LCP index, B = HCP index) associated with the meteorites in (c).

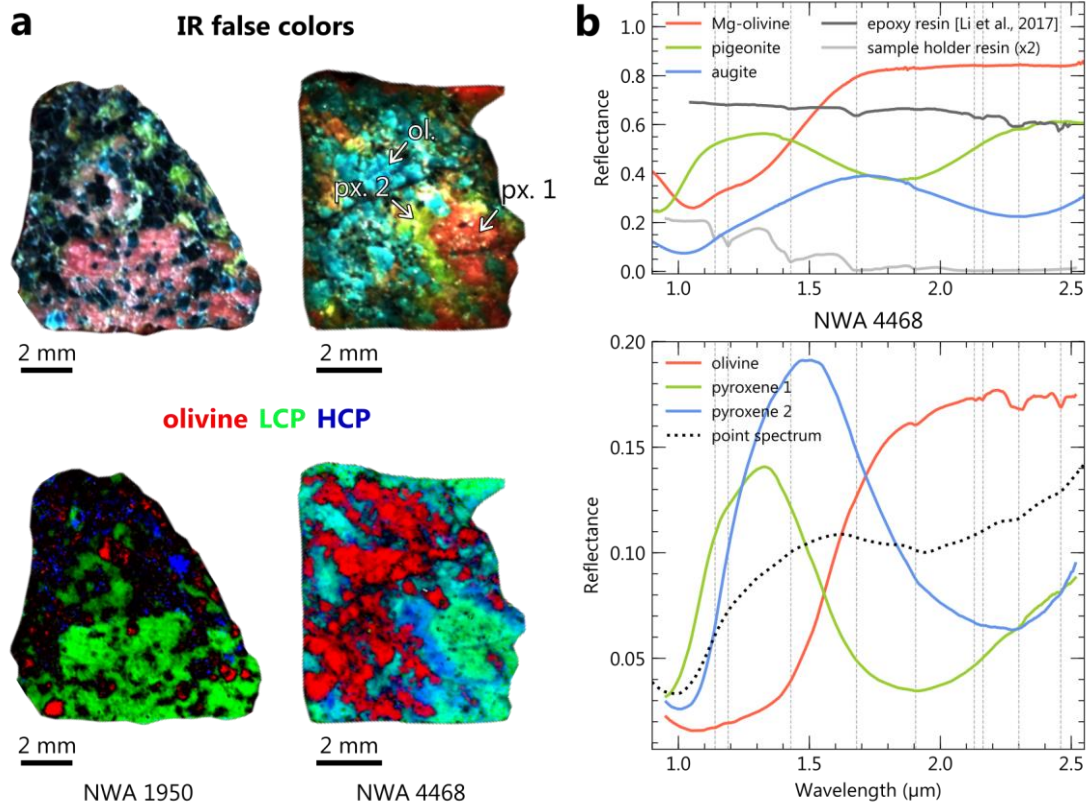

**Figure 7.** Hyperspectral cube of poikilitic shergottites. **(a)** False RGB color composite in the near-infrared (R = 1.22 µm, G = 1.66 µm, B = 2.15 µm) and associated RGB spectral parameters map below (R = olivine index, G = LCP index, B = HCP index). In the IR false color composite chosen, olivine, LCP and HCP appear blue, red/pink and green/yellow respectively. **(b)** Spectra extracted from the hyperspectral cubes of NWA 1950 and NWA 4468 (solid lines, location of spectral extraction shown with arrows on panel (a)) and point spectra obtained with the point spectrometer on a large part of the sample surface (dotted lines, see Fig. 1 for measurement location), compared with laboratory reflectance spectra of known minerals taken from the PDS Geoscience Spectral Library. Laboratory spectra IDs are as follows: forsterite C3PO51, pigeonite C1PP42 and augite C1SC35. The silver resin spectrum was measured outside of the meteorite, in the resin sample holder. Dashed lines refer to the absorption bands of another type of resin. Olivine, LCP and HCP are detected in the sample spectra, with overprinting of two types of resin signature.

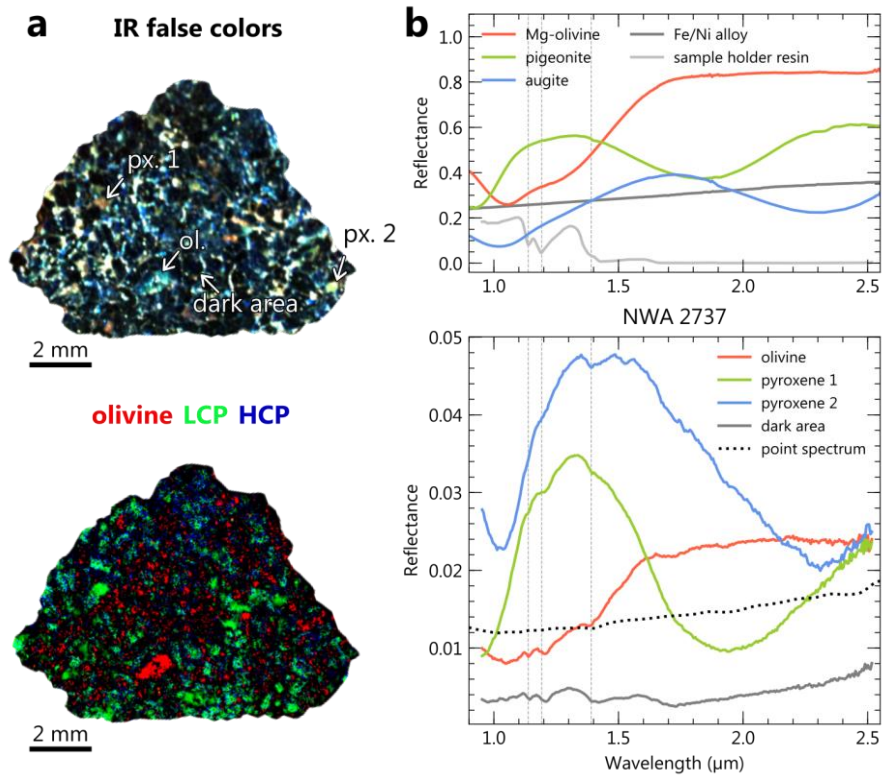

**Figure 8.** Hyperspectral cube of chassignite NWA 2737. **(a)** False RGB color composite in the near-infrared (R = 1.22 µm, G = 1.66 µm, B = 2.15 µm) and associated RGB spectral parameters map below (R = olivine index, G = LCP index, B = HCP index). In the IR false color composite chosen, olivine, LCP and HCP appear blue, red/pink and green/yellow respectively. **(b)** Spectra extracted from the hyperspectral cube of NWA 2737 (solid lines, location of spectral extraction shown with arrows on panel (a)) and point spectra obtained with the point spectrometer on a large part of the sample surface (dotted lines, see Fig. 1 for measurement location), compared with laboratory reflectance spectra of known minerals taken from the PDS Geoscience Spectral Library. Laboratory spectra IDs are as follows: forsterite C3PO51, pigeonite C1PP42, augite C1SC35 and Fe/Ni alloy C1SC12. Resin spectrum was measured outside of the meteorite, in the resin sample holder. Dashed lines refer to the absorption bands of resin. Olivine, LCP and HCP are detected in the sample spectra, with overprinting of the resin signature. The dark areas correspond to olivine crystals. Their reflectance signature (i.e., low reflectance values and a moderate red slope over the VNIR range) is consistent with the presence of Fe/Ni alloy previously reported in this sample (e.g., Pieters et al., 2008).

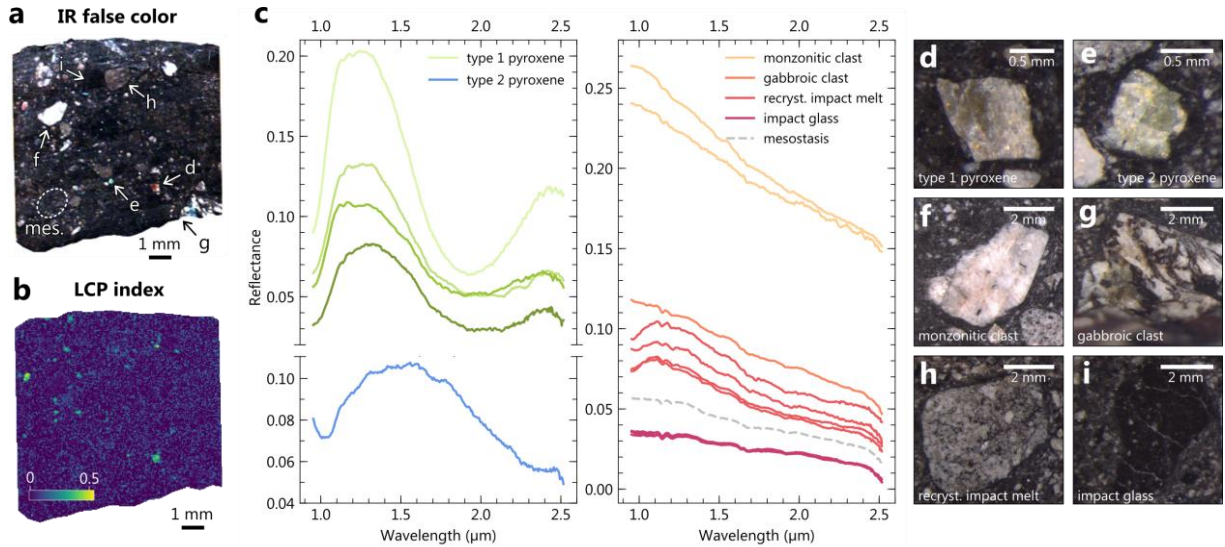

**Figure 9.** Hyperspectral cube of the polymict breccia NWA 7034. **(a)** False RGB color composite in the near-infrared (R = 1.22 μm, G = 1.66 μm, B = 2.15 μm). In the IR false color composite chosen, LCP and HCP appear red/pink and green/yellow respectively. **(b)** Associated LCP spectral parameter map. **(c)** Spectra extracted from the hyperspectral cube of NWA 7034. Pyroxene types are defined by their spectral signature. Associated types of clasts observed through a stereo microscope are shown in **(d, e, f, g, h, i)**. Apart from pyroxenes, the clasts in NWA 7034 are associated with a blue slope in the near-infrared range, similarly to the point measurements on large clasts and over the whole chip. Recrystallized impact melts are associated with a subtle absorption band at ~1 μm, in agreement with the presence of small embedded pyroxene crystals.

Most of the spectra acquired on the polymict breccia NWA 7034 are dominated by a blue spectral slope and are not exhibiting any absorption band in the near-infrared (Fig. 9). Spectra of various types of clasts big enough to be resolved are extracted: pyroxene crystals (Fig. 9d and e), monzonitic feldspathic clasts (Fig. 9f), a gabbroic clast (Fig. 9g), pockets of crystalline and grainy pyroxene-feldspar impact melt (Fig. 9h), and pockets of dark impact glass with embedded small crystals (Fig. 9i). While no clear spectral signature of pyroxene is isolated with point spectroscopy due to spot size limitation (see section 4.2), the majority of the pyroxene clasts have spectra consistent with LCP, and one clast consistent with HCP in the hyperspectral cube (Fig. 9b and c). Isolated felsic clasts show high reflectance levels while not exhibiting any particular absorption band (Fig. 9c). The mesostasis, dark impact glass and gabbroic clasts do not exhibit any clear absorption band, while the pyroxene-feldspar crystalized impact melt pockets show the subtle absorption of pyroxene at ~1 μm (Fig. 9c).

In the NWA 4468 cube, shallow absorptions at 1.91 μm, 2.31 μm and 2.45 μm are detected in association with olivine (Fig. 7). The same bands are observed for NWA 1950 in association with olivine grains and in association with some LCP in areas (Fig. 8). While a mixture with a phyllosilicate or carbonate phase is not excluded as they are likely to produce absorptions at 1.9 μm and 2.3 μm, the set of shallow bands observed are consistent with epoxy contaminants (Fig. 7 and Fig. 8). In the DaG 489 meteorite (not mounted with resin), bright areas show absorptions at 1.39 μm, 1.92 μm, 2.31 μm and 2.40 µm, consistent with the presence of a Mg-phyllosilicate, the best match being a saponite smectite (Fig. 6). While a hydrous carbonate can present absorptions at ~1.4 µm, ~1.9 µm and ~2.3 µm, the distinctive 2.4 µm absorption identified in DaG 489 spectra is typical of phyllosilicate, not carbonate. In NWA 1068 (2) sample, the absorptions at ~1.4 µm and ~1.9 µm are associated with the bright veins crosscutting the rock and melt pockets and veins, but no additional band useful for mineralogical characterization are observed. No other shallow absorption bands are clearly identified in NWA 1068 (1), NWA 2737, SaU 008 and Tissint cubes.

**4.2 Results from point spectra**

*4.2.1 Absolute reflectance*

Martian meteorites are mostly dark in the VNIR range. At the standard observation geometry ($\phi$ = 30°, a = 0°), the mean maximum reflectance of chips/cut samples is 15%, and 95% of them do not have a reflectance exceeding 25%. Such low reflectance values are expected for magmatic rocks with low-silica content (e.g., Sgavetti et al., 2006). Powders are unsurprisingly more reflective with a mean maximum reflectance of 34%, the most reflective sample being NWA 4766 (the maximum observed reflectance is 57%). The darkest rocks from our suite are the nakhlite CeC 022 and the augite basalt NWA 8159, absorbing respectively at least 97% and 99% of the light in the explored spectral range. Finally, powders exhibit redder spectra than rock samples (e.g., NWA 4766 in Fig. 10a).

*4.2.2 Band parameters related to main minerals*

Except for the polymict breccia NWA 7034, the chassignite NWA 2737 and the augite-basalt NWA 8159, Martian meteorites reflectance spectra exhibit two broad absorption bands centered around ~1 µm and ~2 µm (Fig. 10): bands I and II. These bands are explained by the combination of the absorption bands of the main mafic minerals present in the meteorites: pyroxene and olivine (Fig. 2). In addition, some spectra from the Martian meteorite suite exhibit shallow and narrow absorption bands at ~1.9 µm and ~2.3 µm, sometimes associated with an absorption at ~1.4 µm and ~2.45 µm (Fig. 11).

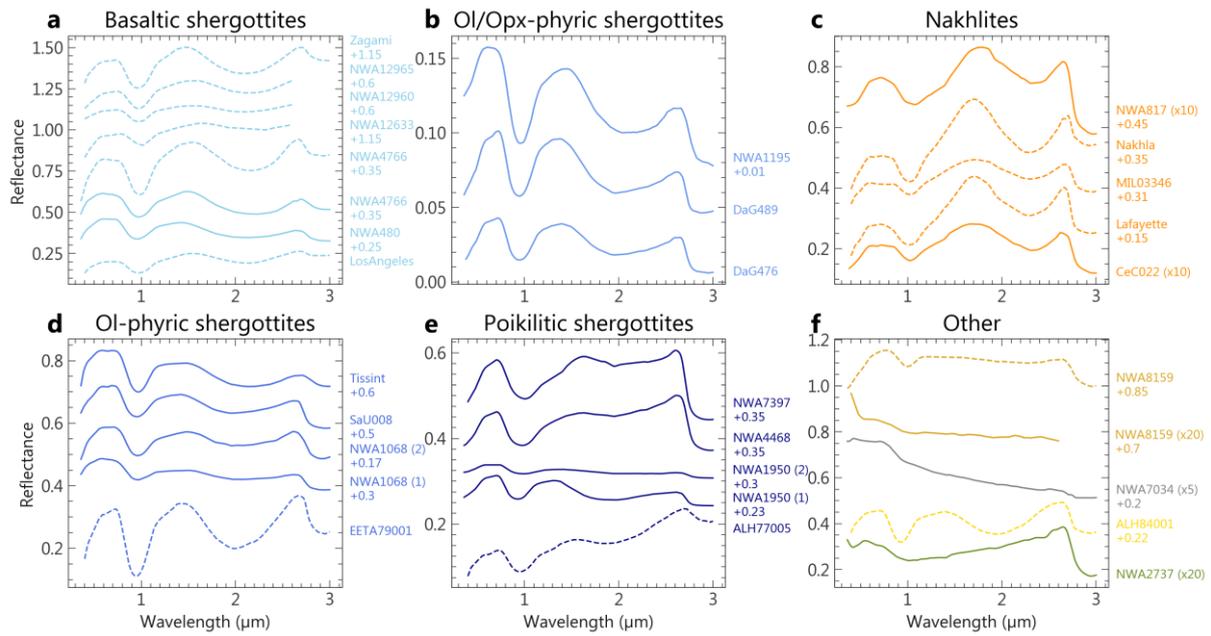

**Figure 10.** Spectra of the Martian meteorite suite obtained with the point spectrometer SHADOWS. Dashed and solid spectra respectively correspond to powder and chip/cut section samples. The spectra of CeC 022, NWA 817, NWA 2737, NWA 7034 and NWA 8159 are amplified in reflectance for clarity. The location of spot measurements on chip/cut section samples is given in Fig. 1.

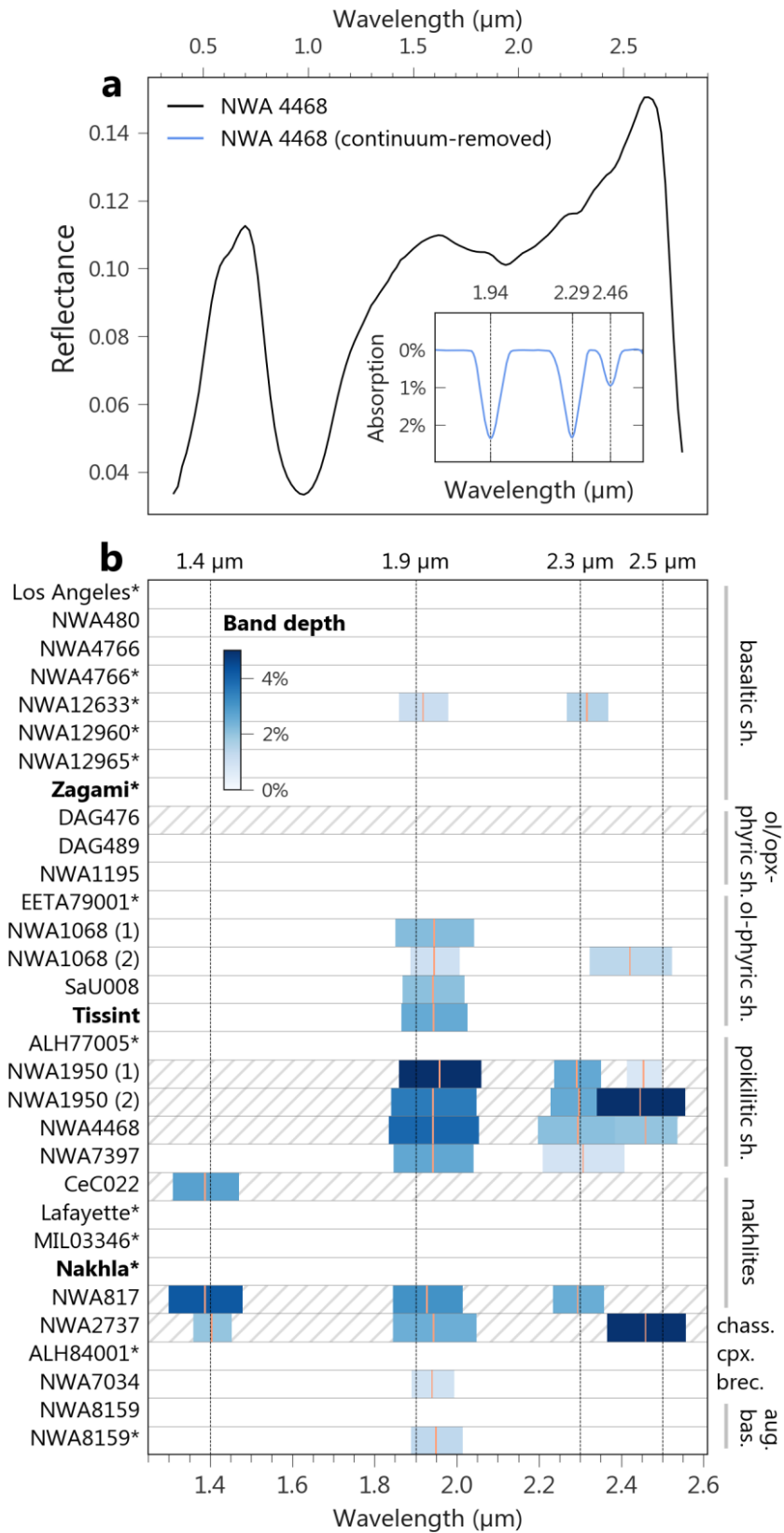

**Figure 11. (a)** Point spectrum of shergottite NWA 4468 showing evidence of absorption bands at 1.95 μm, 2.30 μm and 2.45 μm. The continuum-removed spectrum is obtained by fitting and removing the 2 μm band of pyroxene. **(b)** Position and strength of the narrow absorption bands present in the meteorite point spectrum in the near-

infrared range. The band centers are indicated in blue. Falls are shown in bold and powders are indicated by an asterisk sign. Samples mounted on epoxy (which might present absorptions in this wavelength range) are shown with hatching.

When mixed with pyroxene (usually the dominant phase in Martian meteorites), plagioclase or maskelynite can be indirectly detected by a decrease of the pyroxene band depths and an increase of the absolute value of reflectance at the band shoulders (Crown and Pieters, 1987). In our dataset, flattening of the reflectance peak of pyroxene between band I and band II is observed in the shergottite spectra, resulting in a particularly broad band I (e.g., NWA 1068 (2); Fig. 10). A similar feature was commonly observed on previously measured shergottite spectra; however, applications of the Modified Gaussian Method (MGM) by Sunshine et al. (1993) and McFadden and Cline (2005) showed that the combination of low and high-calcium pyroxenes spectra solely could account for this effect. Hence, flattening of the spectral shoulder of pyroxene in our dataset cannot be used as a means to infer a plagioclase/maskelynite spectral detection. In addition, no correlation between the documented proportion of maskelynite in the shergottites and the reflectance of the pyroxene shoulders or the depth of the absorption is observed.

We report the position, depth, width and FWHM of band I and II of the meteorites' spectra in Figure 12. The absorption position centers are compared with the ones measured on various natural pyroxenes by Adams et al. (1974). Consistently with the prevalent mineralogy of the meteorites, shergottites and nakhlites align well on the trend of clinopyroxenes, whereas ALH 84001 is closer to orthopyroxenes (Fig. 12a).

Nakhlites are well distinguished from shergottites in the position of band I versus position of band II diagram (Fig. 12a): position of band I is higher than ~1.02 µm and position of band II higher than ~2.25 µm. This is in line with the pyroxene Ca content: nakhlites dominant mineral, augite, is Ca-rich, hence the absorptions centered at longer wavelengths. Nakhlites spectra are also associated with specific band widths: band I is usually wider than 0.8 µm, which is the case for some shergottites. This is in line with their petrology: in nakhlites, pyroxene is primarily augite, which exhibits a band I wider than the other pyroxene types (Fig. 2). However, band II is narrower for nakhlites than for other meteorites, with a width smaller than ~1 µm, and a FWHM smaller than ~0.6 µm (Fig. 12). We did not find absorptions of alteration minerals in the spectra of nakhlites.

In shergottite spectra, the position of band I is lower than ~1 µm and the position of band II lower than ~2.2 µm (Fig. 12a). Band I is usually slightly narrower than in the case of nakhlites (narrower than 0.98 µm), while band II is generally broader (larger than 0.98 µm, with a FWHM superior to ~0.57 µm). The different types of shergottites have distinct spectral characteristics, though they are subtler than between nakhlites and shergottites. Basaltic shergottites fall on the trend of natural pyroxenes from Adams et al. (1974) and have band I and band II centered after ~0.95 µm and ~2.07 µm respectively (Fig. 12a). Shergottites with olivine are not perfectly aligned on the pyroxene trend, with the position of band I over the position of band II ratio as well as the band I depth over band II depth II ratio increasing with the olivine content (Fig. 12a and 12c). This is clearly seen in the poikilitic ("lherzolitic") shergottites, which have the highest olivine content and a deeper band I, whose center is closer to their band II, compared to other shergottites (Fig. 12a). Spectra of olivine-bearing shergottites display band I at longer wavelengths, which is consistent with the relative position of the 1 µm band of olivine and the band I of low-Ca pyroxene. Olivine and olivine/orthopyroxene-phyric shergottites have a position of band I over position of band II ratio intermediate between those of poikilitic shergottites and natural pyroxenes, and share a band I and a band II centered below ~0.95 µm and ~2.1 µm respectively (Fig. 12a). These two subgroups are hardly distinguished from one another based on their band I and II. The augite basalt NWA 8159 exhibits faint absorptions at 0.98 µm (intermediate between those of shergottites and nakhlites) as well at 2.28 µm (in the nakhlites values range).

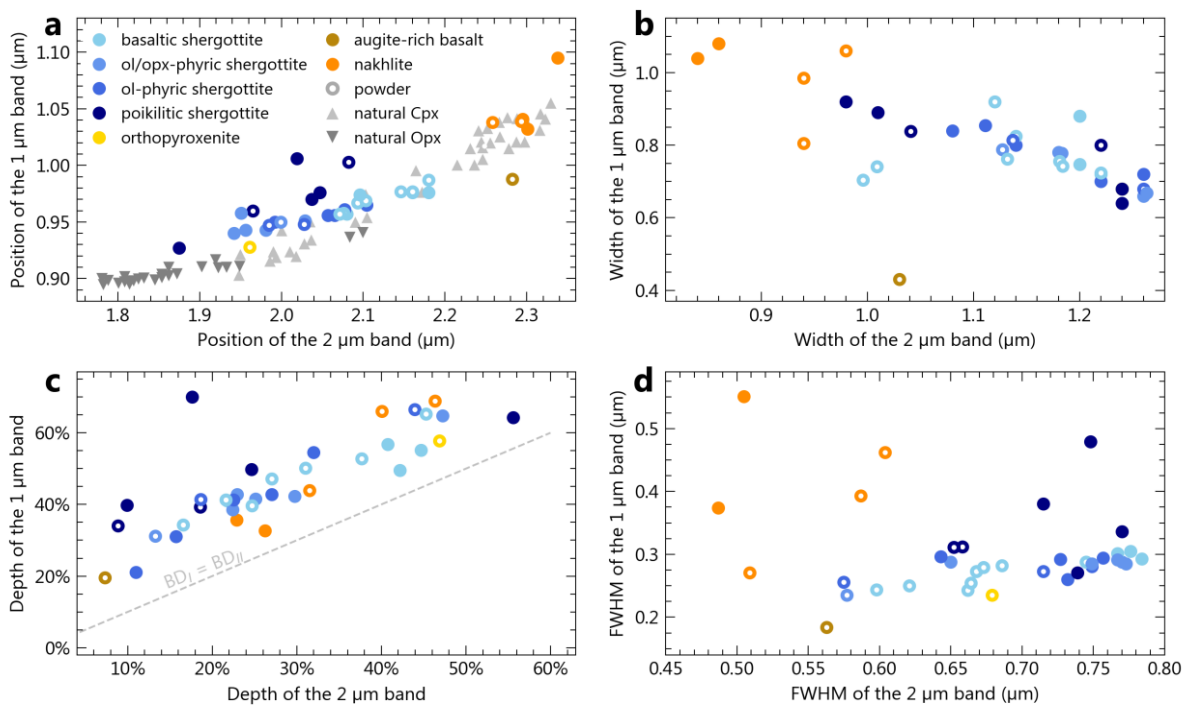

**Figure 12.** Parameters of the meteorite spectra corresponding to the two absorption bands of pyroxene around 1 µm and 2 µm. Natural pyroxenes were measured by Adams et al. (1974). Other Martian meteorites whose absorptions at 1 µm and 2 µm were measured in previous studies were added to the plots: QUE* 94201, NWA 6963, Shergotty (Filiberto et al., 2018), DaG 670, NWA 2626, NWA 6234, NWA 6963 and Y† 984028 (RELAB spectral library; all of these samples are shergottites).

* Queen Alexandra Range

† Yamato

### 4.2.3 Discrimination between single pyroxene and multiple pyroxenes-bearing rocks

Based solely on the absorption band positions, a pyroxene of intermediate calcium content is spectrally hardly distinguishable from a mixture of low- and high-calcium pyroxenes. For instance, the point spectrum of shergottite NWA 480 shows a band II typical of a pyroxene of intermediate calcium composition, while being actually composed of low and high-calcium pyroxenes (Fig. 13e). It has previously been proposed that band width could be used to aid in the determination between pure pyroxene and pyroxene mixture (e.g., Klima et al., 2007), as the superposition of LCP and HCP spectra would result in wider absorptions. Figure 13 shows a comparison between band width and FWHM of pure pyroxene spectra and spectra of measured meteorites – both on meteorites with multiple pyroxene mineralogy (i.e., LCP and HCP) and those dominated by a single pyroxene mineralogy (i.e., mostly LCP or HCP). These sets of measurements align together on the same trends and are hardly distinguishable based on the width or FWHM of the pyroxene absorption bands. While not discriminant, FWHM of the 2 µm band seems to be the best parameter to help distinguish mixture of pyroxenes and single-pyroxene mineralogy, as most of synthetic pyroxenes exhibit a slightly narrower band II (Fig. 13d). Interestingly, multiple pyroxenes meteorites spectra usually exhibit a band I narrower than those of synthetic pyroxenes, which is unexpected (Fig. 13a).

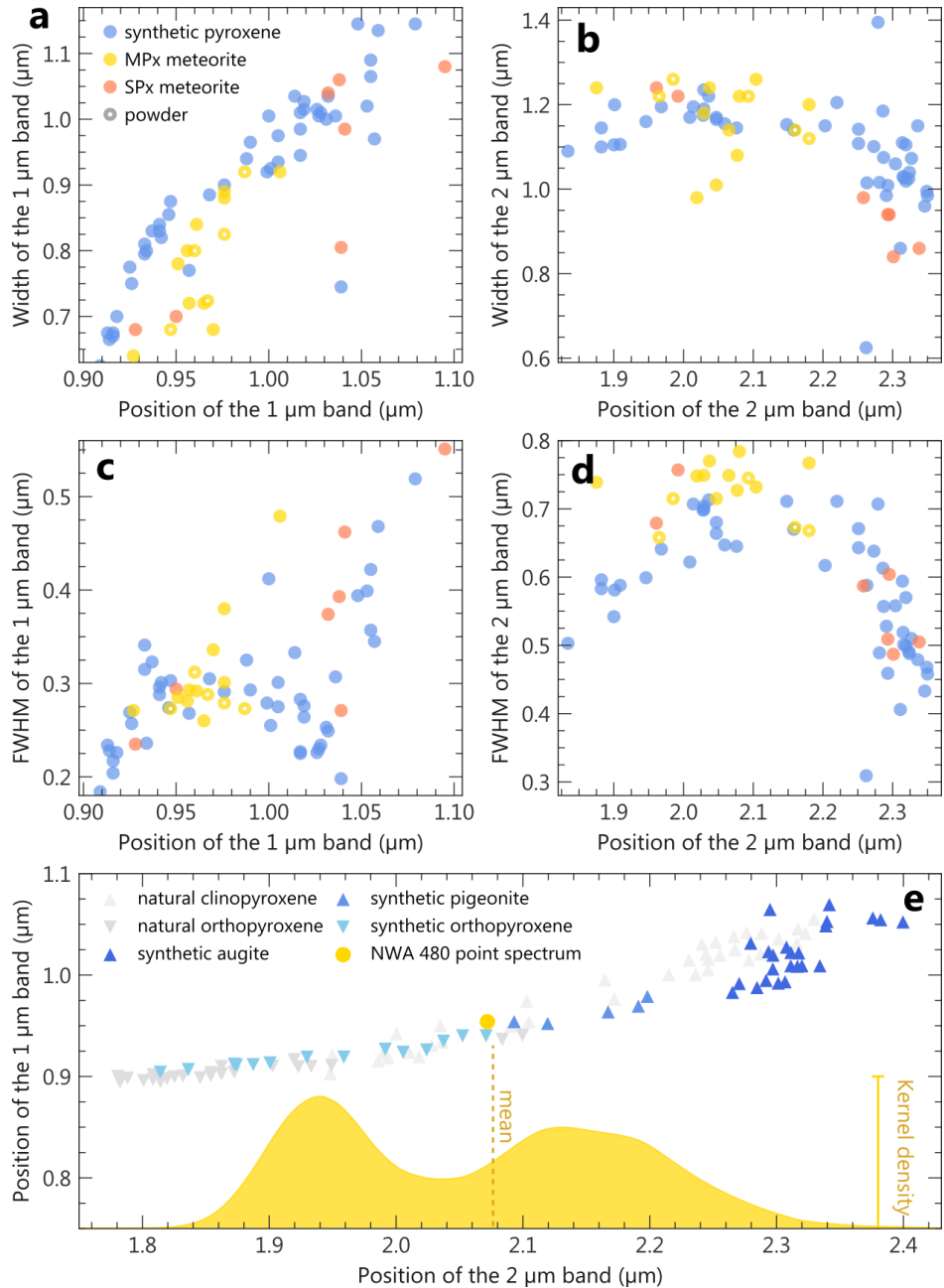

**Figure 13. (a, b, c, d)** Width and FWHM of the bands I and II as a function of band position for synthetic pyroxenes (Klima et al., 2007; 2011), compared to those of the point spectra of meteorites measured in this study. MPx refers to meteorites with multiple pyroxene mineralogy (i.e., LCP and HCP) while SPx refers to meteorites dominated by a single pyroxene mineralogy (i.e., mostly LCP or HCP). **(e)** Band positions of NWA 480 point spectrum and cube spectra, compared to those of synthetic (Klima et al., 2007; 2011) and natural (Adams et al., 1974) pyroxenes spectra.

*4.2.3 NWA 7034, the poymict breccia*

Point spectra acquired on a chip of NWA 7034 (polymict breccia) are consistent with previous VNIR measurements (Beck et al., 2015; Cannon et al., 2015) and show similarities with spectra from the hyperspectral cube: the sample is darker than most of the Martian meteorites from our suite, with a maximum reflectance of 11% in the visible and 7% at 2.5 μm (Fig. 10). This has been interpreted to be caused by the elevated content in magnetite and maghemite oxides (Beck et al., 2015). Whole chip and clasts spectra display a notable blue slope in the near-infrared (Fig. 14). Varying spectral slope as well as varying strength of the 3 µm hydration absorption band are observed from one clast to another. However, out of the seven clasts investigated, and apart from the 3 µm band, only one clast exhibits an absorption band in this spectral range (Fig. 14), showing that most of the breccia has spectral features dominated by the presence of oxides. This shallow band located at 0.95 µm is likely caused by the presence of a LCP clast in the measurement spot (Fig. 2). Limitation on the beam size prevented the analysis of smaller clasts, but further details are provided by imaging spectroscopy (section 4.1).

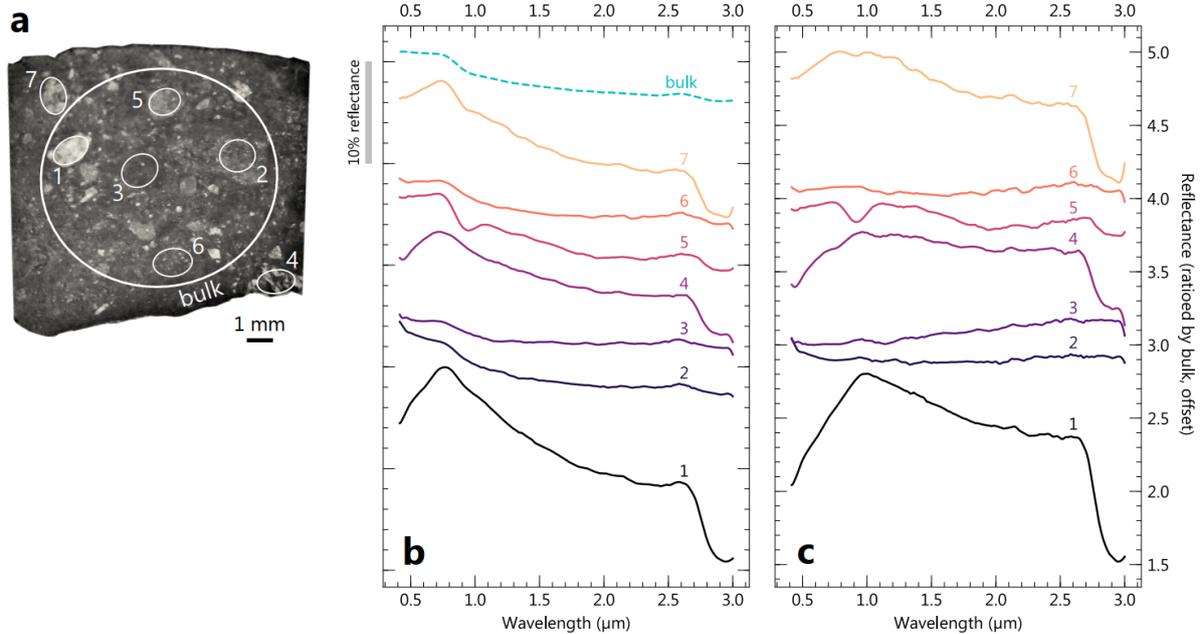

**Figure 14.** Reflectance spectra of NWA 7034 (polymict breccia). **(a)** Location of spot measurements on chip sample. **(b)** Spectra of various clasts are compared to the spectrum acquired over the whole cut section ("bulk"). **(c)** The same reflectance spectra are ratioed by the bulk spectrum.

### 4.3 Influence of the observation geometry

Bidirectional spectra acquired at various observational geometries on the particulate sample and cut section of NWA 4766 are shown in Figure 15. For both samples, variations of absolute reflectance and bands strength with changing observational geometries are observed.

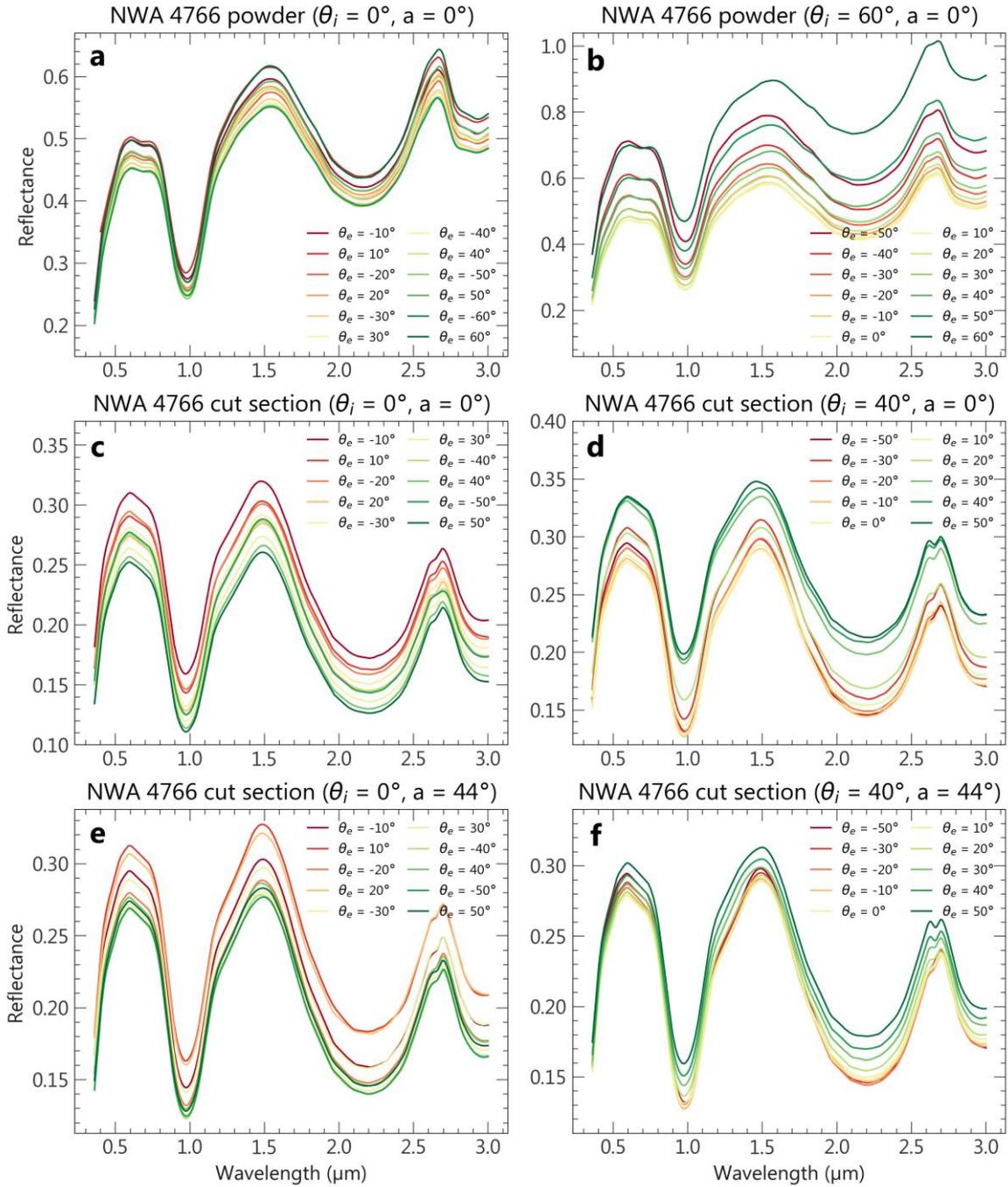

**Figure 15.** Spectra of NWA 4766 (basaltic shergottite) acquired with the spectro-gonio radiometer SHADOWS at various observational geometries, with emergence angles between -60° and 60° by steps of 10° for the powder and between -50° and 50° by steps of 10° for the cut section sample. Spectra are color-coded based on the absolute value of the phase angle (low: red, high: green). Technical limitations restrict measurements at low phase angles, hence spectra corresponding the following emergence angles are absent: **(a)** 0°, **(b)** -60°, **(c)** 0°, **(d)** -40°, **(e)** 0°, **(f)** -40°.

*4.3.1 Strength of the reflected light for varying emergence/phase and incidence angle*

The lowest reflectance levels are usually observed for phase angles of 40-70°, while the highest are observed at phase angles close to 0°, as well as at the highest ones we were able to measure (90° for the cut section and 120° for the powder; Fig. 16). Overall, both particulate and rock samples behave as non-Lambertian surfaces and show some increase of their absolute reflectance at low phase angle, indicative of backward scattering of the light by the sample. This effect is best depicted in BRDF (Bidirectional Reflectance Distribution Function) polar diagrams, where the intensity of the reflected rays at a given wavelength (here outside of the absorption bands of pyroxene) is shown as a function of their emergence angle $\theta_e$ (Fig. 17). Since high phase angle measurements were limited, there are no data points at $\phi$ higher than 90° and 120° for the cut section and powder, respectively. This means that any forward scattering behavior is hardly explored through our measurements. However, measurements at $\phi > 90°$ on the powder show an increase of the reflected light intensity towards higher phase angle, indicating that the sample does produce some forward scattering. The intensity of the backward scattering, which we measured by ratioing the reflectance at $\theta_i = 0°$, $\theta_e = 30°$ by the reflectance at the lowest phase angle measurement ($\theta_i = 0°$, $\theta_e = 10°$), is relatively similar for the particulate sample and for the cut section (Fig. 16). At 600 nm, these ratios are equal to 1.07 and 1.10 for the powder and cut section respectively. The same intensity of forward scattering is observed for measurements at ($\theta_i = 0°$, $\theta_e = -10°$) and ($\theta_i = 0°$, $\theta_e = -30°$), despite the asymmetry in the reflectance level patterns (Fig. 16).

Continuum reddening of the powder spectra is seen to increase with the phase angle, at any incidence. For instance, the reflectance at 2.68 μm at $\theta_i = 60°$ and $\theta_e = 60°$ ($\phi = 120°$) is ~1.4 times the reflectance at 0.60 μm, while it is ~1.1 times at $\theta_i = 60°$ and $\theta_e = -50°$ ($\phi = 10°$). This effect is also observed on the cut section spectra but is minor compared to the powder observed at the same geometry, with up to 5% of reddening at $\phi = 90°$ compared to $\phi = 10°$.

At low phase angle, the intensity of the light reflected by the powder increases for oblique incident light (i.e., high $\theta_i$) compared to geometries where the incoming light source is at nadir (i.e., $\theta_i = 0°$; Fig. 16 and Fig. 17), meaning that the sample tends to produce more backscattered light when illuminated obliquely.

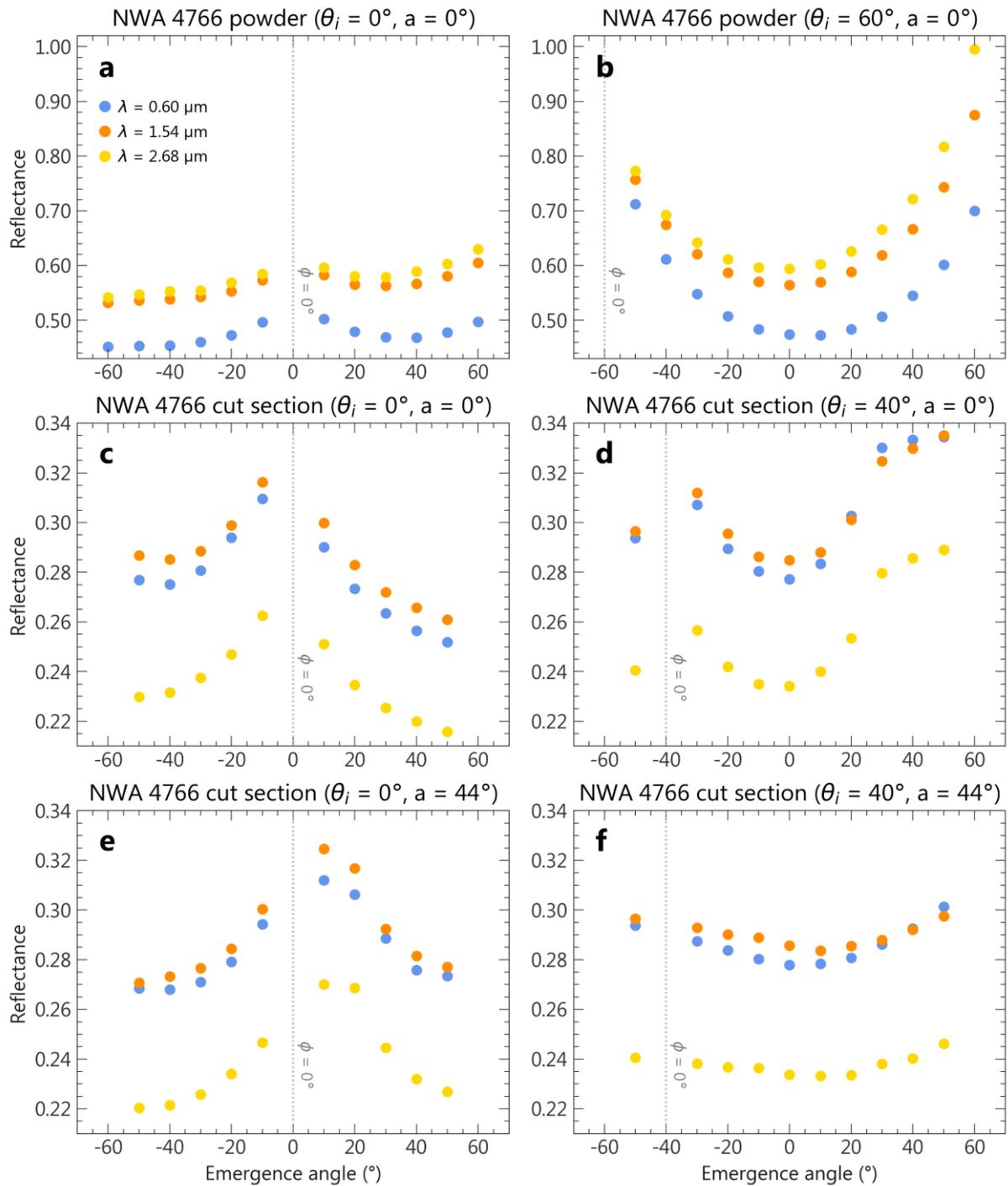

**Figure 16.** Reflectance of NWA 4766 (basaltic shergottite) obtained at various incidence, emergence and azimuth angles, with emergence angles between -60° and 60° by steps of 10° for the powder and between -50° and 50° by steps of 10° for the cut section sample. The reflectance was measured outside of the absorption bands of pyroxene, at the following wavelengths: 0.60 μm (blue dots), 1.54 μm (orange dots), 2.68 μm (golden dots). Asymmetry in the reflectance levels pattern is explained by non-perfectly flat samples during measurement.

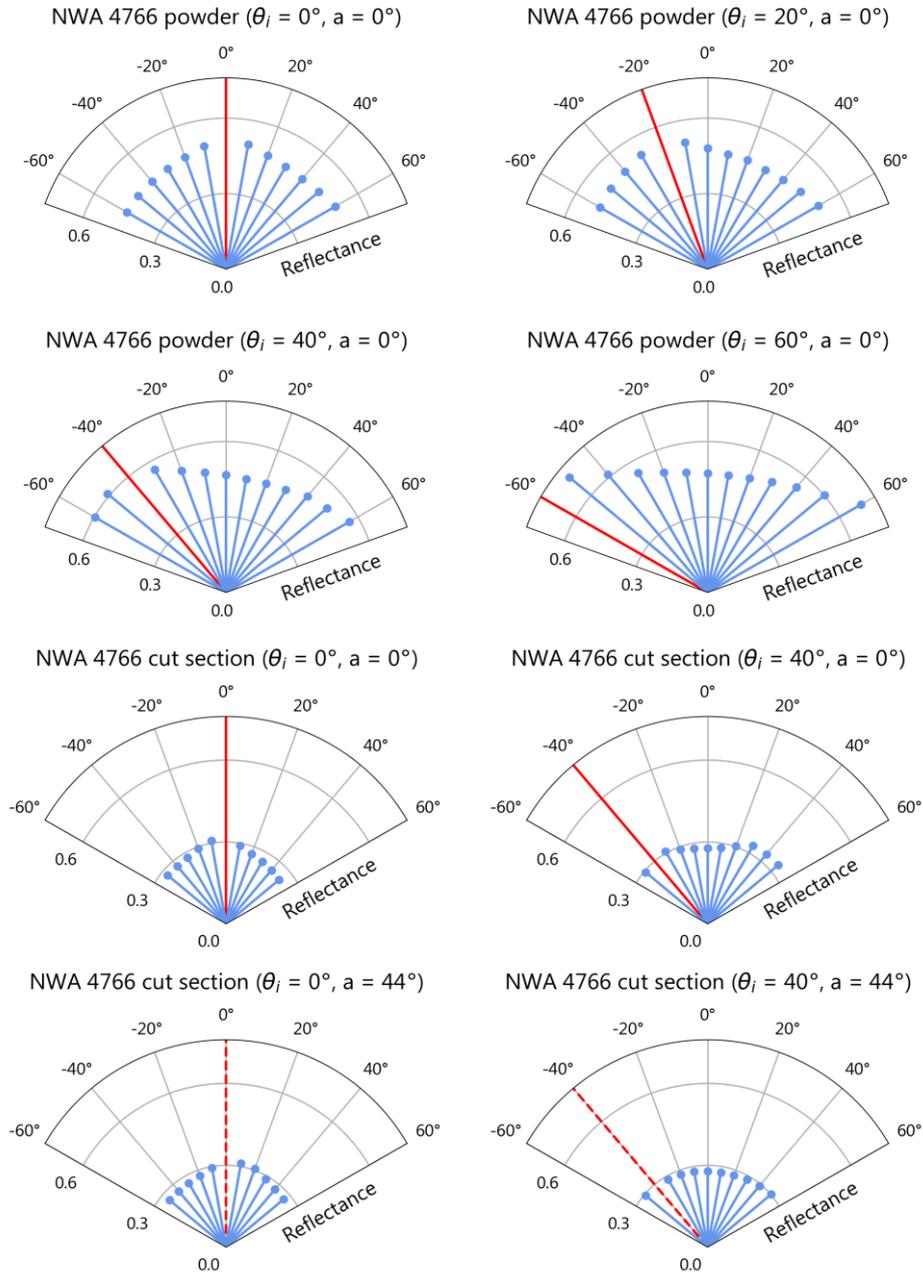

**Figure 17**. BRDF at 0.6 µm for NWA 4766. The blue lines correspond to the reflectance for various emergence angles (see Table 1 for angle values). The incidence angle is indicated with a red line, dashed when the illumination is not in the principal plane ($a \neq 0°$). Moderate backward scattering behavior is observed for both types of sample, while only the particulate sample shows clear moderate forward scattering behavior (at the observation geometries explored).

*4.3.2 Reflected light strength outside of the principal plane*

The effects of measuring the reflected light at varying azimuth angle on reflectance spectra can be assessed through bidirectional measurements on the cut sample, where the reflected light was measured at positive azimuth (Table 1). Observations with variable azimuth for vertical illumination seem to generate similar spectra in terms of absolute reflectance (see Fig. 16c compared to 16e) and band depths (see red curve compared to orange curve in Fig. 18d and 18e) than measurements in the principal plane. At oblique incident light, changes in azimuth produce distinct spectra than those measured in the principal plane: the absolute reflectance shows little variation at diverse emergence angles (Fig. 15, Fig. 16, and Fig. 17), suggesting that forward scattering probably does not extend laterally outside of the principal plane.

Overall, considering all the bidirectional spectra at various incidence, emergence and azimuth angles, the maximal variation of reflectance observed at the same wavelength is 9% for the cut section and 46% for the powder. This value for the powder drops to 14% when considering only the observation geometries that were explored commonly for both sample types (the variation for the cut sample remains unchanged). If the two types of samples and all geometries are considered, the maximum reflectance variation spectrum-to-spectrum is up to 79% of reflectance for NWA 4766.

*4.3.3 Effect of the observational geometry on band parameters*

No significant shift in band positions is observed with varying observational geometry: no more than 5 nm for band I and no more than 60 nm for band II. These little variations appear uncorrelated from the reddening of the spectra and are most likely caused by uncertainties in continuum removal, or slight variations of the grains measured between different observational geometries (as the measurement spot variates).

The strength of band I and II show variations with changing observation angles, respectively up to 12% and 9% for the particulate sample and up to 15% and 13% for the cut section. This leads to the $BD_I/BD_{II}$ ratio increasing with the phase angle in most cases, with up to 10% of variation given the observational geometry. The variation of band depths depending on the observational geometry seems to be anticorrelated with the variation of reflectance: maximum band depths for both absorption bands are observed when the reflectance is low ($\varphi \sim 40\text{-}60°$), while the lowest band depths are observed at $\varphi$ close to 0° and at $\varphi = 120°$ which correspond to high absolute reflectance (Fig. 16 and 17). Similar to what is observed for the absolute reflectance, band depths appear to fluctuate less at high incidence and high azimuth (green curve in Fig. 18d and 18e), than in the principal plane at the same incidence (yellow curve in Fig. 18d and 18e).

Considering both samples and all explored geometries, strength of band I and II vary by up to 17% and 23% respectively.

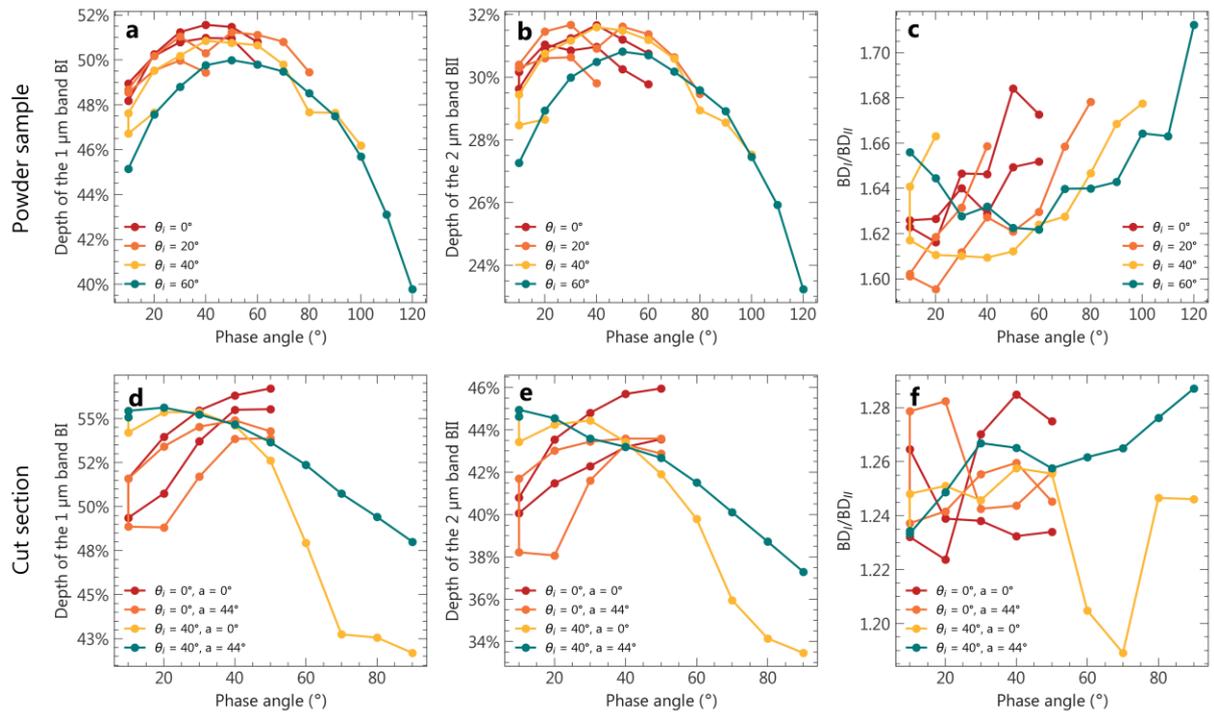

**Figure 18.** Variation of the pyroxene bands parameters as a function of the observational geometry, for the powder **(a, b, c)** and cut section **(d, e, f)** samples of NWA 4766.

## 5 Discussion

### 5.1 Comparison of point spectra and cubes

As expected, the point spectra acquired on the meteorites with the SHADOWS spectrometer result from the spectral mixing of various components. In NWA 480, three main phases are identified in the spectral imagery: the 2 µm absorption band centers indicative of low and high-calcium pyroxenes (from the core to the rim of the crystals), and a groundmass with an intermediate 2 µm band center and lower reflectance values. Comparison with SHADOWS measurement shows that these three components are spectrally mixed in the point spectrum of NWA 480, which exhibits intermediate 2 µm band center and moderate reflectance values (Fig. 5b). In ol-phyric and ol/opx-phyric shergottites cubes, olivine phenocrysts and dark areas embedded in a pyroxene groundmass with a dominant low-calcium signature are identified, while their point spectra are dominated by the spectral signature of the groundmass (Fig. 6b). Low reflectance value in the point spectra can be explained by the presence of the olivine phenocryst and the dark areas, which have lower reflectance in the hyperspectral cubes. The point spectra of the poikilitic shergottites also appear to be a spectral mixing of olivine and pyroxene signatures, with reflectance values intermediate between those of the olivine, LCP and HCP crystals (Fig. 7b).

These results indicate that significant spectral mixing occurs in the point measurements at spot sizes down to a few millimeters, which is SuperCam's spot size at a few meters from the target (Wiens et al., 2017). At this resolution, imaging spectroscopy (i.e., MicrOmega's technique) is more powerful to isolate the various main minerals and minor phases. However, our previous results on band parameters show that point spectrometry is sufficient to discriminate the different Martian meteorites families and to identify their main mineralogy.

**5.2 Nature of the alteration in Martian meteorites**

Alteration phases have been reported in many of the meteorites studied here (see dedicated section 2.2). We show that some Martian meteorites have spectra with shallow absorption bands in the near-infrared, which usually point to the presence of alteration phases. In particular, bands at ~1.9 μm, ~2.3 μm and ~2.4-2.5 μm, usually attributed to hydrated minerals are observed (Fig. 11). Unfortunately, resin contaminants likely present in several of the samples and absorbing at these same wavelengths prevent any definitive conclusion on the ability of reflectance spectrometry to identify these phases in most of our Martian meteorites suite. Nonetheless, several other samples exhibit these bands while never have been embedded in resin: DaG 489, NWA 1068 pairs, NWA 7034, NWA 7397, NWA 8159, NWA 12633, SaU 008 and Tissint (Fig. 6 and 11).

Our measurements with imaging spectroscopy show that, in DaG 489, a smectite phase occurring in patches in the sample is the cause of the shallow absorption bands observed in the point spectrum. In paired DaG 476, the bright mineral phase occurring in fractures, in grain boundaries and filling cavities was identified to be a Ca-carbonate (Greshake and Stoeffler, 1999; Zipfel et al., 2000; Mikouchi et al., 2001). While carbonate, if present in our sample, is not detected in DaG 489 with imaging spectroscopy, Greshake and Stoeffler (1999) and Mikouchi et al. (2001) reported iddingsite (mixture of oxides, ferrihydrites and poorly crystalline phyllosilicates rich in iron and magnesium) associated with altered olivine grains of DaG 476. The iddingsite is probably the phase detected here, and in particular the phyllosilicate component, which we identify to be a Mg-smectite (a common product of olivine alteration, e.g., Dehouck et al., 2014).

Similar to the Dar al Gani pair, the terrestrial Ca-carbonate phase which fills fractures in the NWA 1068 pair (Barrat et al., 2002) was not detected in the point spectra or in the hyperspectral cubes. Instead, these white fractures have the spectral signature of hydration, with bands at ~1.4 μm and ~1.9 μm, and lack the carbonate bands at ~2.3 μm and ~2.5 μm. Hydrous carbonates can have similar features with subdued 2.3 μm and 2.5 μm (Harner et al., 2015) and, although more common in cold desert finds, were reported in hot desert finds (Miyamoto, 1991). However, hydrous Ca-carbonate such as monohydrocalcite are usually metastable with respect to calcite and aragonite (Hull and Turnbull, 1973) and are not expected to precipitate extensively in a hot desert environment, or to be preserved during a long-term storage at room temperature. Hydrous Mg-carbonates such as hydromagnesite or nesquehonite on the other hand, can be stable if the formation of magnesite (anhydrous Mg-carbonate) is kinetically inhibited (i.e., by temperature or $CO_2$ partial pressure), and have been reported in weathered chondrites (Marvin and Motylewski, 1980; Hänchen et al., 2008). Consequently, a terrestrial hydrous Mg-carbonate might be the phase detected in the NWA 1068 pair.

The alteration phases causing the shallow absorptions in the other meteorites without resin are quite challenging to determine, as hyperspectral imaging of the samples fails at isolating these phases to retrieve interpretable spectra, at least at the spatial resolution and SNR of our measurements. Of particular interest is the 1.9 µm band identified in Tissint spectrum; as this meteorite is an observed fall, any hydrated mineral is likely of Martian origin. The only mineral that can exhibit hydration and that has been documented in Tissint is apatite, which is reported in minor phases (less than 1%; Chennaoui Aoudjehane et al., 2012). The absorption in Tissint spectrum is very subtle (~1% of absorption after removal of the pyroxene 2 µm band), which is agreement with the very low content of hydrated minerals reported. Also, water adsorption during the measurement cannot be excluded.

**5.3 Observational geometry considerations**

As expected, the Martian sample studied both in powder and section form (NWA 4766) shows a non-Lambertian behavior with significant backward scattering. A forward scattering peak is observed for the powder sample, while it could not be investigated on the chip sample (as no measurement at a phase angle higher than 90° was performed).

Notably, an anticorrelation between reflectance and band depth was measured: at observational geometries minimizing the reflectance level (usually for $\phi$ ~40—60°), maximization of band depth was observed, and vice versa. Pommerol and Schmitt (2008), Shepard and Cloutis (2011) and Beck et al. (2012) observed the same trend at high phase angle, respectively on a smectite and volcanic tuff, on lazurite and on meteorites. Pommerol and Schmitt (2008) proposed that, for these observational geometries, the proportion of photons scattered by direct reflections on the surface grains compared to the absorbed photons is more important, causing a decrease of band strength at high phase angle. This arch shape (Fig. 18) observed for the evolution of band depths as a function of phase angle was observed by Beck et al. (2012) and Schröder et al. (2014) with reflectance ratios and is likely linked with the spectral slope, though band depths formulation is more complex than reflectance ratios.

The increase of reddening with phase angle in the powder's spectra was also observed by Beck et al. (2012) and Schröder et al. (2014), where some microscopic roughness effects, unpredicted by the classical radiative models, are proposed to explain this spectral behavior.

The SuperCam visible and infrared spectrometers acquire spectra at different sol hours, which makes the phase angle and azimuth between the Sun (light source) and the rover's Mast Unit (which collects the reflected light) highly variable from one observation to another (Fouchet et al., submitted). Previous bidirectional measurements have already shown the significant impact of a changing observational geometry on the spectral reflectance levels and band strengths. Hence, the main goal of SuperCam reflectance spectrometers will be the mineralogical identification rather than the quantification, which can be biased if relying on band depth. We quantified the significance of these changes on a Martian sample for phase angles ranging from 10° to 120° and for positive azimuth. Though the rocks that will be analyzed by the Perseverance rover will be very diverse, and will include sedimentary and igneous rocks (e.g., Goudge et al., 2015; 2017; Horgan et al., 2020), our results illustrate the expected variations in reflectance and band depths due solely to changes in the observation geometry.

For our sample and angles explored, the absolute reflectance of the powder can be doubled, while variations of only ~10% are observed for the rock sample. Thus, while caution should be applied when interpreting future measurements from SuperCam for the reasons developed above, it is noted that the effects of variable observation geometries on the absolute reflectance are less important for the rock than for the loose material. In addition, while the powder is remarkably more reflective than the rock slab, pyroxene band strengths are comparable in spectra of both samples. Of interest for the identification of subtle bands associated to alteration minerals, there is a band strength optimum depending on the phase angle (in our case between 40° and 60°). However, this optimum also corresponds to minimal absolute reflectance, hence lower SNR measurements. A phase angle compromise between absolute reflectance and band strength is recommended. Measurements performed at angle ±ϕ around this optimum are expected to be comparable both in terms of reflectance level and band depth. Additionally, measurements performed with different azimuths are also expected to be comparable in terms of absolute reflectance and band depths if they are acquired close to the nadir – if ever achieved on Mars.

**5.4 Implications for the spectral study of igneous rocks on Mars**

Our sample suite reflects the current global diversity of Martian meteorites, where effusive rocks are rare and intrusive rocks dominant. Thus, the spectral features of most of our samples should not be directly compared to the orbital observations of the surface – except from outcrops where rocks have been excavated from depth, by erosional processes or during an impact (e.g., rocks in central peaks of complex craters; Brustel et al., 2019). Nevertheless, this case study can be interpreted in terms of phase detectability and mixing in the spectral data.

Over the last decades, reflectance spectroscopy between ~1 μm and ~2.6 μm has been used to study the diversity of the Martian crust at a global scale (e.g., Bibring et al., 2005; Mustard et al., 2005). The predominant detection of pyroxene (and olivine) indicates that the crust is dominated by poorly evolved lithologies. The detection of felsic rocks by reflectance and thermal infrared measurements over the highlands (Christensen et al., 2005; Bandfield et al., 2006; Carter and Poulet, 2013; Wray et al., 2013), in situ at Gale crater (Sautter et al., 2015) and in the polymict breccia NWA 7034 and pairings (Agee et al., 2013) showed that at least locally, evolved lithologies exist on Mars. While feldspars are more challenging to detect than pyroxenes when mixed with other minerals (Crown and Pieters, 1987), our measurements on meteorites at high resolution show that, even isolated from the pyroxenes (i.e., in spectral cubes), they are hardly detectable. High shock pressure during the impact(s) is likely to decrease the ~1.3 μm absorption band depth of feldspars (Johnson and Hörz, 2003) but does not alone explain the complete absence of absorption in our spectral data. Other explanations on the non-detection of this band include low-iron content (as the absorption is caused by $Fe^{2+}$) or the presence of darkening impurities in the feldspars. Felsic clasts in NWA 7034 are broadly featureless in the VNIR range: they sample a primitive Martian crust that is probably undetectable from the orbit with reflectance spectroscopy.

Important interpretations about the magmatic evolution and cooling history of the Martian crust are derived from the types of pyroxene detected from the orbit (e.g., Baratoux et al., 2013). We show that, based on spectral parameters, Martian meteorites with both low- and high-calcium pyroxene are hardly distinguishable from rocks with a single intermediate type of pyroxene, while these rocks can be formed via different magmatic processes (e.g., the presence of exsolution lamellae of new composition can occur with slow cooling). We highlight the necessity to consider spectral mixtures when interpreting pyroxene-dominated spectra, where deconvolution methods like the Modified Gaussian Models (MGM) should be applied (Sunshine and Pieters, 1993). Nonlinear unmixing approaches such as the Hapke (Hapke, 1981) or Shkruratov (Shkruratov, 1999) models might also address this issue, but they were developed for particulate materials where grains are separated. Their applicability to compact rocks remains to be assessed and validated.

## 6 Summary

By collecting an extensive suite of samples representative of the current Martian meteorites diversity, we were able to produce spectral measurements of several samples of each family in the VNIR range. The key findings are summarized below.

- The measurement spot of the spectro-gonio radiometer SHADOWS is analogous to the size of those of Mars 2020 SuperCam VNIR spectrometer at ~2–4 m distance (Wiens et al., 2017). Overall, our measurements with SHADOWS are able to recover the primary mineralogy of the Martian meteorites suite, based on the absorption bands parameters: olivine, low and high-calcium pyroxene. The presence of maskelynited plagioclase is not clearly detected. Although not abundant in Martian meteorites, some hydrous phases are detected, but their identification is impeded by the spot size and is eventually achieved using high-resolution imaging spectroscopy (MicrOmega's technique). The alteration phases identified likely consist of a hydrous Mg-carbonate of terrestrial origin (in NWA 1068) and a Mg-smectite (in DaG 489).
- The most effective distinction within Martian meteorites based on the point spectra is achieved using the positions of band I versus position of band II criterion, which enables the identification of nakhlites over shergottites, having different pyroxene composition and olivine contents. Similarly, further discrimination can be made inside the various shergottite classes, between basaltic, phyric and poikilitic shergottites. No correlation between spectral properties and ejection ages is observed.
- At 2–4 m from the outcrop, the point spectra acquired by SuperCam are likely to suffer from spectral mixing. Here, and as confirmed by imaging spectroscopy, the point spectra acquired by SHADOWS are the result of a mixing between – when present – high and low-calcium pyroxenes, olivine and additional darkening phases (e.g., melt, oxides or alloys). Rocks dominated by a single type of pyroxene with intermediate calcium content and rocks dominated by both LCP and HCP cannot be easily distinguished from each other based on band parameters. However, using the FWHM of band II seems to help in the discrimination, being more often higher in the case of mixtures.

- Multiple bidirectional measurements on a shergottite, the most abundant type of Martian meteorites, confirm their non-Lambertian behavior in both particulate and consolidated form, with forward scattering (at least for the powder) and backward scattering of the VNIR light. While strongly variable reflectance levels are observed for the powder with varying observational geometry, these variations are lower for the consolidated sample. The pyroxene absorption strengths show variations up to ~10–15%, comparable in both types of sample.
- Of interest for in situ identification of minor minerals, the absorption strengths are maximal at moderate phase angle (here between 40° and 60°). Measurements performed outside of the principal plane and/or at angle ±ϕ around this optimum are expected to be comparable both in terms of reflectance level and band depth.

The meteorite spectra are available for further studies in the supplementary materials.

## Acknowledgments

We thank B. Reynard, J.-A. Barrat and F. Moynier for providing a significant number of meteorite samples analyzed in this study. L. Mandon and C. Quantin-Nataf have been supported by the Agence Nationale de la Recherche (ANR, ANR-18-ERC1-0005) and by the Centre National d'Études Spatiales (CNES) in France. P. Beck has been supported by the H2020 European Research Council (ERC, SOLARYS ERC-CoG2017_771691).